%% file: main.tex
\documentclass{article}
\usepackage{amsmath}
\usepackage{graphicx}
\usepackage{xcolor}

\usepackage{xr}
\makeatletter
\newcommand*{\addFileDependency}[1]{
  \typeout{(#1)}
  \@addtofilelist{#1}
  \IfFileExists{#1}{}{\typeout{No file #1.}}
}
\makeatother

\usepackage{nameref}
\usepackage[labelfont=bf]{caption}
\usepackage{subcaption}

\usepackage{float}
\fontfamily{cmss}\selectfont
\graphicspath{{figures/}}
\usepackage{soul}
\usepackage{newfloat}
\DeclareFloatingEnvironment[name={Figure},fileext=lsf,listname={List of Supplementary Figures}]{suppfigure}

\usepackage[pagebackref=true,breaklinks=true,colorlinks,bookmarks=false]{hyperref}

\title{A Number Sense as an Emergent Property \\ of the Manipulating Brain}

\author{*N. Kondapaneni and P. Perona \\
California Institute of Technology
}
\date{\today}

\begin{document}
\maketitle

\section{Abstract}
The ability to understand and manipulate numbers and quantities emerges during childhood, but the mechanism through which humans acquire and develop this ability is still poorly understood. We explore this question through a model, assuming that the learner is able to pick up and place small objects from, and to, locations of its choosing, and will spontaneously engage in such undirected manipulation. We further assume that the learner's visual system will monitor the changing arrangements of objects in the scene and will learn to predict the effects of each action by comparing perception with a supervisory signal from the motor system. We model perception using standard deep networks for feature extraction and classification, and gradient descent learning. Our main finding is that, from learning the task of action prediction, an unexpected image representation emerges exhibiting regularities that foreshadow the perception and representation of numbers and quantity. These include distinct categories for zero and the first few natural numbers, a strict ordering of the numbers, and a one-dimensional signal that correlates with numerical quantity. As a result, our model acquires the ability to estimate {\em numerosity}, i.e. the number of objects in the scene, as well as {\em subitization}, i.e. the ability to recognize at a glance the exact number of objects in small scenes. Remarkably, subitization and numerosity estimation extrapolate to scenes containing many objects, far beyond the three objects used during training. We conclude that important aspects of a facility with numbers and quantities may be learned with supervision from a simple pre-training task. Our observations suggest that cross-modal learning is a powerful learning mechanism that may be harnessed in artificial intelligence.

\section{Introduction}

\subsection{Background}
Mathematics, one of the most distinctive expressions of human intelligence, is founded on the ability to reason about abstract entities. We are interested in the question of how humans develop an intuitive facility with numbers and quantities, and how they come to recognize numbers as an abstract property of sets of objects.  There is wide agreement that innate mechanisms play a strong role in developing a {\em number sense}~\cite{xu2005number,dehaene2011number,viswanathan2013neuronal}, that development and learning also play an important role~\cite{dehaene2011number}, that naming numbers is not necessary for the perception of quantities~\cite{gordon2004numerical,pica2004exact}, and a number of brain areas are involved in processing numbers~\cite{dehaene1999sources,harvey2013topographic}. Quantity-tuned units have been described in physiology ~experiments~\cite{viswanathan2013neuronal, nieder2009representation, nieder2016neuronal, doi:10.1073/pnas.2201039119} as well as in computational studies~\cite{stoianov2012emergence, zorzi2018emergentist, 2019numbnasrer,kim2021visual}. 

\subsection{Related Work}
The role of learning in developing abilities that relate to the natural numbers and estimation has been recently explored using computational models. Fang et al.~\cite{fang2018can} trained a recurrent neural network to count sequentially and Sabathiel et al.~\cite{sabathiel2020emerging} showed that a neural network can be trained to anticipate the actions of a teacher on three counting-related tasks -- they find that specific patterns of activity in the network's units correlate with quantities. The ability to perceive {\it numerosity}, i.e. a rough estimate of the number of objects in a set, was explored by Stoianov, Zorzi and Testolin~\cite{stoianov2012emergence,zorzi2018emergentist}, who trained a deep network encoder to efficiently reconstruct patterns composed of dots, and found that the network developed units or ``neurons'' that were coarsely tuned to quantity, and by Nasr et al.~\cite{2019numbnasrer}, who found the same effect in a deep neural network that was trained on visual object classification, an unrelated task. In these models quantity-sensitive units are an emergent property. In a recent study, Kim et al.~\cite{kim2021visual} observed that a random network with no training will exhibit quantity-sensitive units. After identifying these units,~\cite{stoianov2012emergence, zorzi2018emergentist, 2019numbnasrer,kim2021visual} train a supervised classifier on a two-set comparison task to assess numerosity properties encoded by the deep networks. These works showed that training a classifier with supervision, in which the classifier is trained and evaluated on the same task and data distribution, is sufficient for recruiting quantity-tuned units for relative numerosity comparison. Our work focuses on this supervised second stage. Can more be learned with less supervision? We show that a representation for numerosity, that generalizes to several tasks and extrapolates to large quntities, may arise through a simple, supervised pre-training task. In contrast to prior work, our pre-training task only contains scenes with up to 3 objects, and our model generalizes to scenes with up to 30 objects.

\subsection{Approach}
We focus on the interplay of action and perception as a possible avenue for this to happen. More specifically, we explore whether perception, as it is naturally trained during object manipulation, may develop representations that support a number sense. In order to test this hypothesis we propose a model where perception learns how specific actions modify the world. The model shows that perception develops a representation of the scene which, as an emergent property, can enable the ability to perceive numbers and estimate quantities at a glance~\cite{jevons1871power,piazza2002subitizing}.

\begin{figure}
\begin{center} 
    \includegraphics[width=\linewidth]{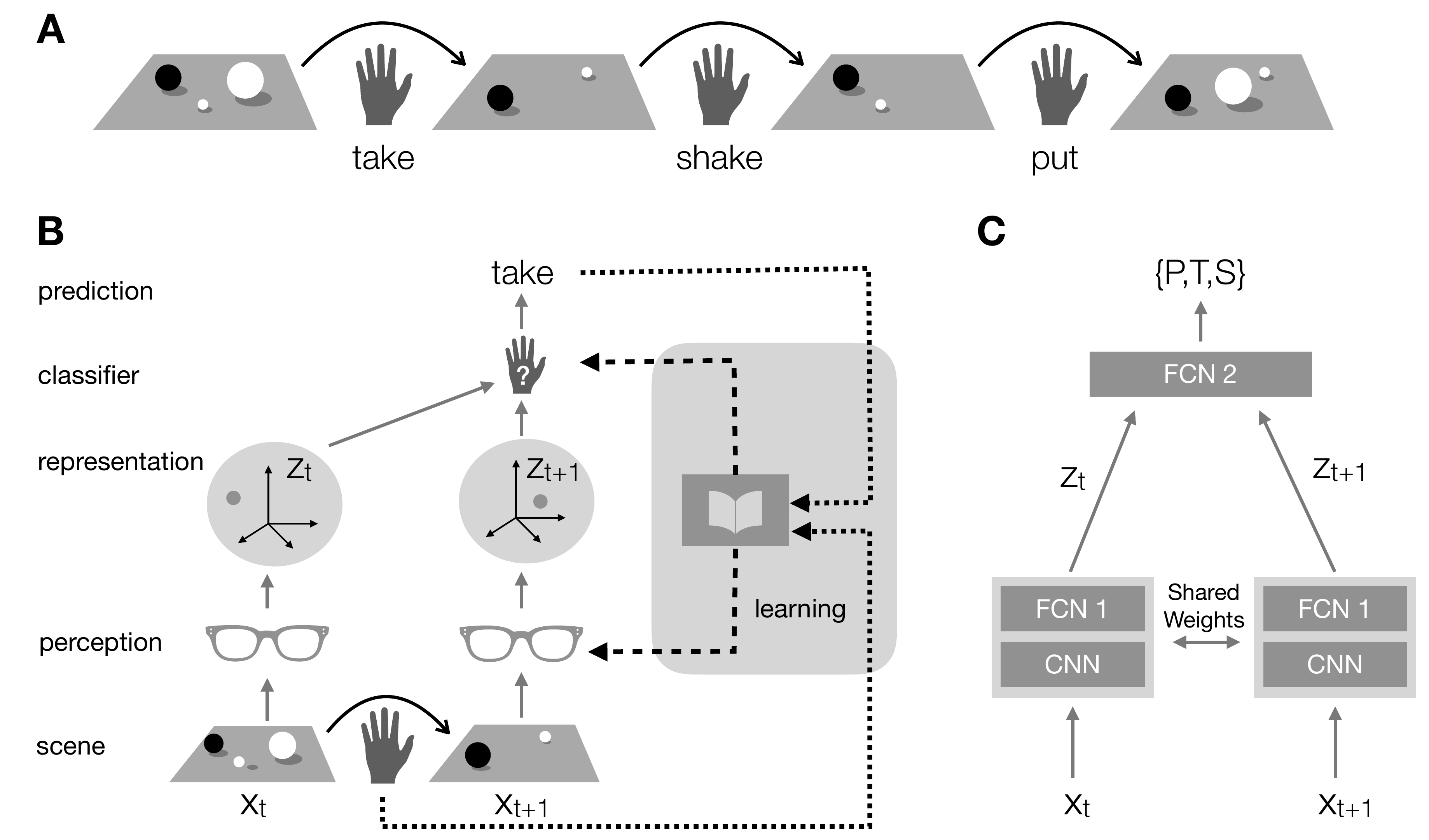}
\end{center}
\linespread{1} \caption{\small {\bf Schematics of our model.} {\bf \textbf{(A)}} (Left-to-right) A sequence of actions modifies the visual scene over time.
{\bf \textbf{(B)}} (Bottom-to-top) The scene changes as a result of manipulation. The images $x_t$ and $x_{t+1}$ of the scene before and after manipulation are mapped by perception into representations $z_t$ and $z_{t+1}$. These are compared by a classifier to predict which action took place. Learning monitors the error between predicted action and a signal from the motor system representing the actual action, and updates simultaneously the weights of both perception and the classifier to increase prediction accuracy. {\bf (C)} (Bottom-to-top) Our model of perception is a hybrid neural network composed of the concatenation of a convolutional neural network (CNN) with a fully-connected network (FCN 1). The classifier is implemented by a fully connected network (FCN 2) which compares the two representations $z_t$ and $z_{t+1}$. The two perception networks are actually the same network operating on distinct images and therefore their parameters are identical and learned simultaneously  in a {\em Siamese network} configuration~\cite{bromley1994signature}. Details of the models are given in Fig.~\ref{fig:network_details}.}
\label{fig:network}
\begin{center}
    \renewcommand{\thesubfigure}{\Alph{subfigure}}
    \includegraphics[width=1\linewidth]{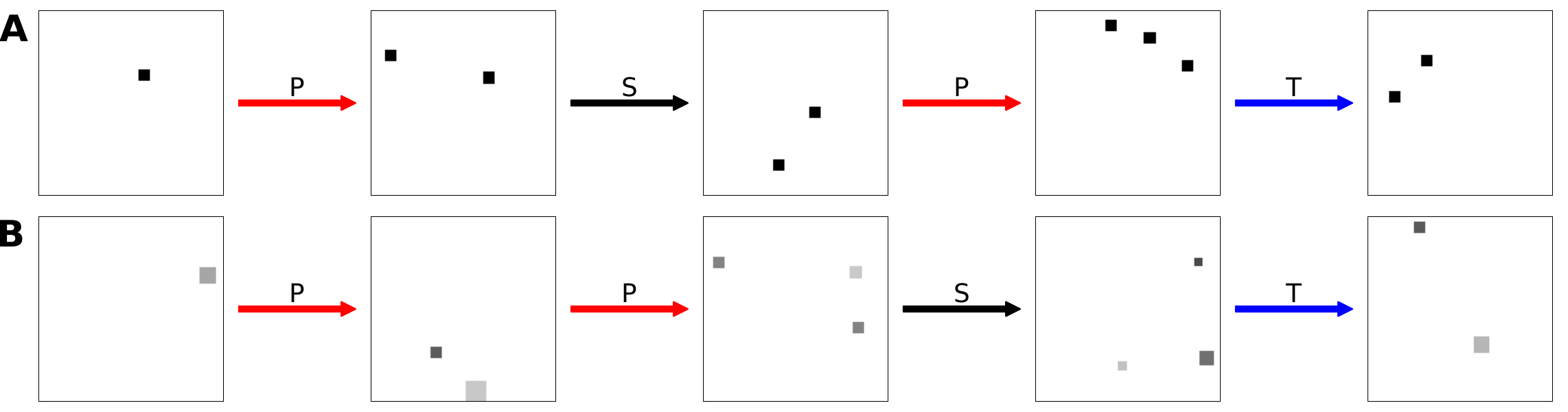}
\end{center}
\linespread{1} \caption{\small {\bf Training image sequence samples.} We trained our model using sequences of images that were generated by randomly concatenating take (T), put (P) and shake (S) manipulations, while limiting the number of objects to the $\left\{ 0 \dots 3 \right\}$ set (see Methods - Training Sets). We experimented with two different environment/scene statistics: {\bf (A)} Identical objects (15x15 pixel squares) with random position. {\bf (B)} Objects (squares) of variable position, size and contrast. The overall image intensity is a poor predictor of cardinality in this dataset (statistics in Fig.~\ref{fig:dataset_statistics}). Images have been inverted to better highlight objects with low contrast.}
\label{fig:main_data}
\end{figure}

In order to ground intuition, consider a child who has learned to pick up objects, one at a time, and let them go at a chosen location. Imagine the child sitting comfortably and playing with small toys (acorns, Legos, sea shells) which may be dropped into a bowl. We will assume that the child has already learned to perform at will, and tell apart, three distinct operations  (Fig.~\ref{fig:network}A). The {\em put} (P) operation consists of picking up an object from the surrounding space and dropping it into the bowl. The {\em take} (T) operation consists in doing the opposite: picking up an object from the bowl and discarding it. The {\em shake} (S) operation consists of agitating the bowl so that the objects inside change their position randomly without falling out. Objects in the bowl may be randomly moved during put and take as well.

We hypothesize that the visual system of the learner is engaged in observing the scene, and its goal is predicting the action that has taken place~\cite{singer2018sensory}as a result of manipulation. By comparing its prediction with a copy of the action signal from the motor system it may correct its perception, and improve the accuracy of its predictions over time. Thus, by performing P, T, and S actions in a random sequence, manipulation generates a sequence of labeled two-set comparisons to learn from.

We assume two trainable modules in the visual system: a ``perception" module that produces a representation of the scene, and a ``classification" module that compares representations and guesses the action (Fig.~\ref{fig:network}). 

During development, perceptual maps emerge, capable of processing various scene properties. These range from basic elements like orientation~\cite{hubel1962receptive} and boundaries~\cite{von1984illusory} to more complex features such as faces~\cite{tsao2006cortical} and objects~\cite{tsao2003faces,hung2005fast}. We propose that, while the child is playing, the visual system is being trained to use one or more such maps to build a representation that facilitates the comparison of the pair of images that are seen before and after a manipulation. These representations are often called embeddings in machine learning.

A classifier network is simultaneously trained to predict the action (P, T, S) from the representation of the pair of images (see Fig.~\ref{fig:network}). As a result, the visual system is progressively trained through spontaneous play to predict (or, more accurately, post-dict) which operation took place that changed the appearance of the bowl.

We postulate that signals from the motor system are available to the visual system and are used as a supervisory signal (Fig.~\ref{fig:network}B). Such signals provide information regarding the three actions of put, take and shake and, accordingly, perception may be trained to predict these three actions. Importantly, no explicit signal indicating the number of objects in the scene is available to the visual system at any time.

Using a simple model of this putative mechanism, we find that the image representation that is being learned for classifying actions, simultaneously learns to represent and perceive the first few natural numbers, to place them in the correct order, from zero to one and beyond, as well as estimate the number of objects in the scene.

We use a standard deep learning model of perception~\cite{lecun1998gradient,krizhevsky2012imagenet,lecun2015deep}:  a {\em feature extraction} stage is followed by a {\em classifier} (Fig.~\ref{fig:network}). The feature extraction stage maps the image $x$ to an internal representation $z$, often called an {\em embedding}. It is implemented by  a deep network~\cite{krizhevsky2012imagenet} composed of convolutional layers (CNN) followed by fully connected layers (FCN 1).  The classifier, implemented with a simple fully connected network (FCN 2), compares the representations $z_t$ and $z_{t+1}$ of the {\em before} and {\em after} images to predict which action took place. Feature extraction and classification are trained jointly by minimizing the prediction error. We find that the embedding dimension makes little difference to the performance of the network (Fig.~\ref{fig:embedding_dimension_fig}). Thus, for ease of visualization, we settled on two dimensions.

We carried out train-test experiments using sequences of synthetic images containing a small number of randomly arranged objects (Fig.~\ref{fig:main_data}).  When training we limited the top number of objects to three (an arbitrary choice), and each pair of subsequent images was consistent with one of the manipulations (put, take, shake). We ran our experiments twice with different object statistics. In the first dataset the objects were identical squares, in the second they had variable size and contrast. In the following we refer to the model trained on the first dataset as \textit{Model A} and the model trained on the second dataset as \textit{Model B}. 

\section{Results}

\begin{figure}
\begin{center}
    \renewcommand{\thesubfigure}{\Alph{subfigure}}
    \includegraphics[width=1\linewidth]{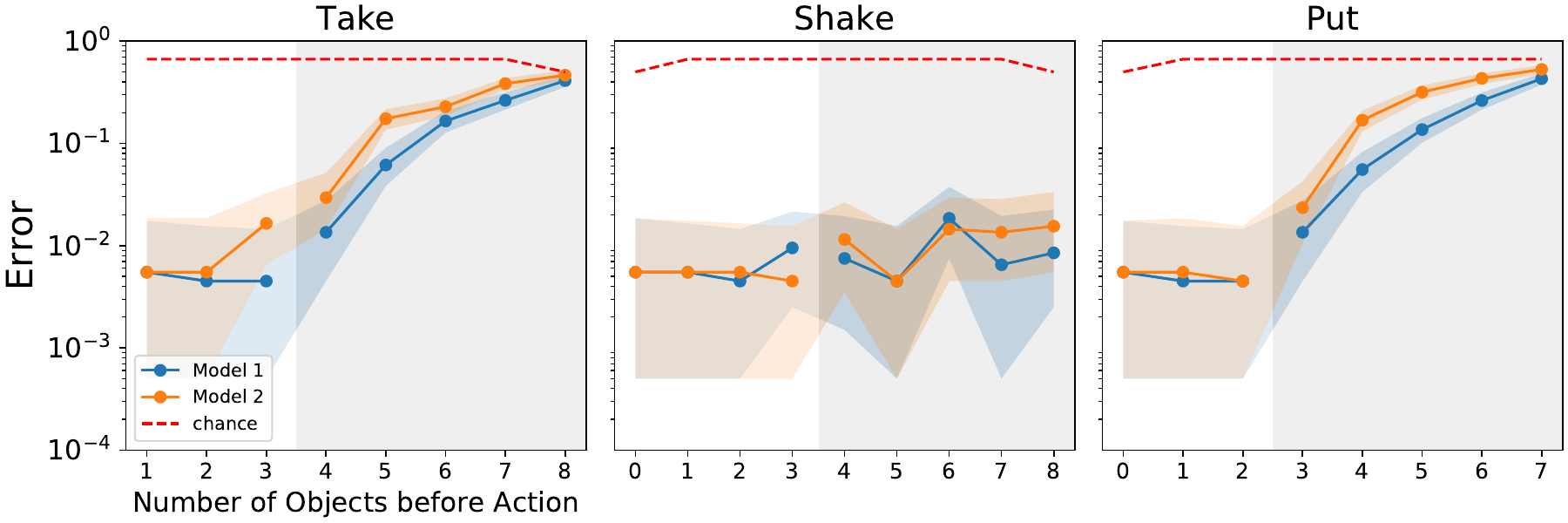}
\end{center}
\linespread{1} \caption{\small {\bf Action classification performance.} The network accurately classifies actions up to the training limit of three objects, regardless of the statistics of the data (the x axis indicates the number of objects in the scene before the action takes place). Error increases  when the number of objects in the test images exceeds the number of objects in the training set. 95\% Bayesian confidence intervals are shown by the shaded areas (272 $\leq$ N $\leq$ 386). The gray region highlights test cases where the number of objects exceeds the number in the training set. The dashed red line indicates chance level.}
\label{fig:errors}
\end{figure}

We found that models learn to predict the three actions on a test set of novel image sequences (Fig.~\ref{fig:errors}) with an error below 1\% on scenes up to three objects (the highest number during training). Performance degrades progressively for higher numbers beyond the training range. Model B's error rate is higher, consistently with the task being harder. Thus, we find that our model learns to predict actions accurately as one would expect from supervised learning. However, there is little ability to generalize the task to scenes containing previously unseen numbers of objects. Inability to generalize is a well-known shortcoming of supervised machine learning and will become relevant later.

\begin{figure}
\begin{center}
    \includegraphics[width=1\linewidth]{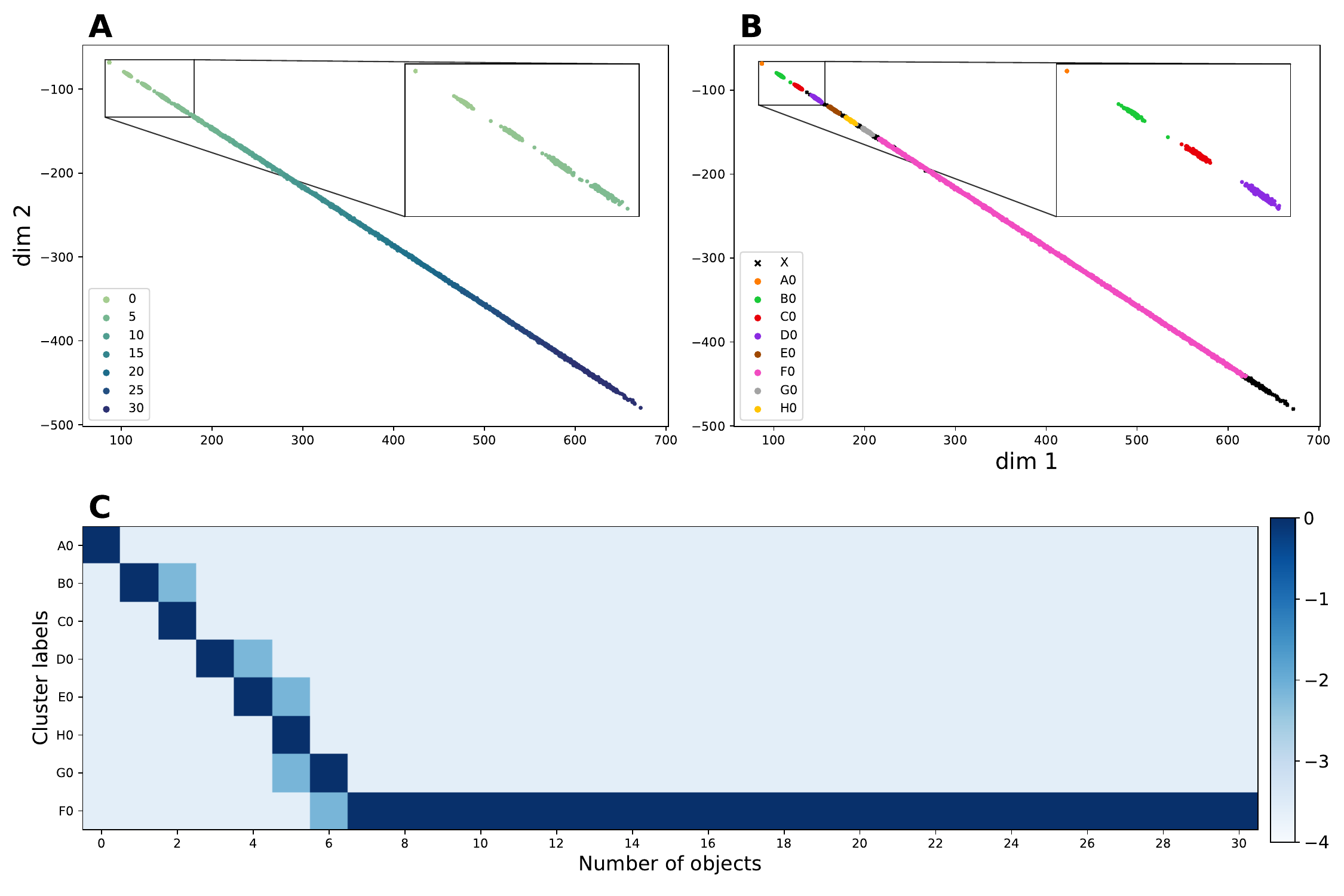}
    \renewcommand{\thesubfigure}{\Alph{subfigure}} 
\end{center}
\linespread{1} \caption{\textbf{The embedding space for Model B.} To explore the structure of the embedding space, we generated a dataset with $\left\{ 0 \dots 30 \right\}$ objects, extending the number of objects far beyond the limit of 3 objects in the training task. Each image in the dataset was passed through Model B and the output (the internal representation/embedding) of the image is shown. See Fig.~\ref{fig:embedding_space_a} for Model A. \textbf{(A)} Each dot indicates an image embedding and the embeddings happen to be arranged along a line. The number of objects in each image is color coded. The smooth gradation of the color suggests that the embeddings are arranged monotonically with respect to the number of objects in the corresponding image. The inset shows that the embeddings of the images that contain only a few objects are arranged along the line into ``islands''. \textbf{(B)} We apply an unsupervised clustering algorithm to the embeddings. Each cluster that is discovered is denoted by a specific color. The cluster X, denoted by black crosses, indicates points that the clustering algorithm excluded as outliers. 
\textbf{(C)} The confusion matrix shows that the clusters that are found by the clustering algorithm correspond to numbers. Images containing 0 - 6 objects are neatly separated into individual clusters; after that images are collected into a large group that is not in one-to-one correspondence with the number of objects in the image. The color scale is logarithmic (base 10).} 
\label{fig:embedding_space}
\end{figure}

When we examined the structure of the embedding we were intrigued to find a number of interesting regularities (Fig.~\ref{fig:embedding_space}). First, the images' representations do not spread across the embedding, filling the available dimensions, as is usually the case. Rather, they are arranged along a one-dimensional structure. This trait is very robust to extrapolation: after training (with up to three objects), we computed the embedding of novel images that contained up to thirty objects and found that the line-like structure persisted (Fig.~\ref{fig:embedding_space}A). This {\it embedding line} is also robust with respect to the dimensions of the embedding -- we tested from two to 256 and observed it each time (Fig.~\ref{fig:embedding_dimension_fig}).

\begin{figure}
\begin{center}
    \includegraphics[width=1\linewidth]{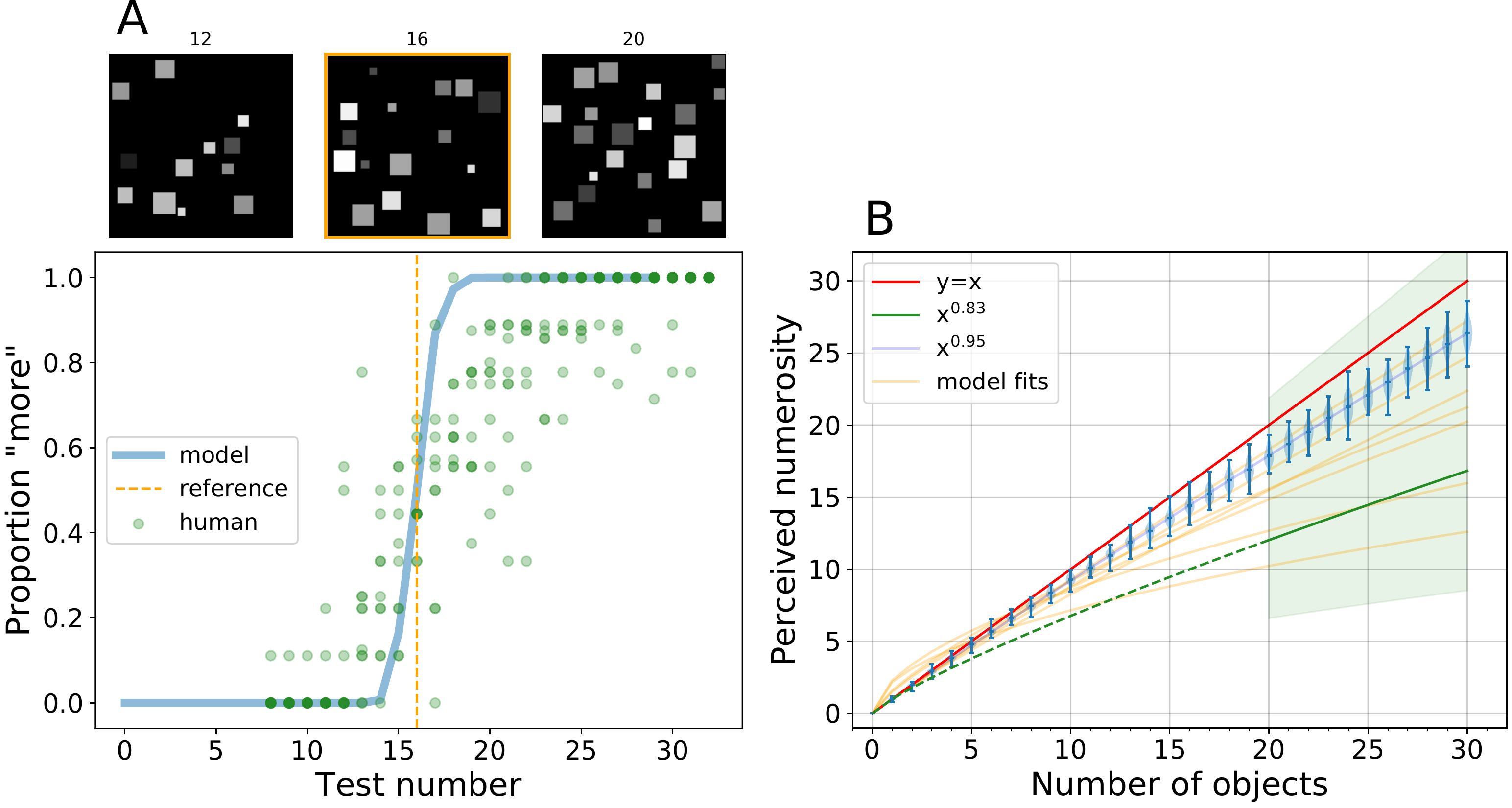}
\end{center}
\linespread{1} \caption{\small \textbf{Relative and absolute estimation of quantity.} \textbf{(A)} Two images may be compared for quantity~\cite{burr2008visual} by computing their embedding and observing their position along our model's embedding line: the image that is furthest along the line is predicted to contain more objects. Here images containing a {\em test number} of objects (see three examples above containing N=12, 16 and 20 objects) are compared with images containing the {\em reference} number of objects (vertical orange dashed line, N=16). The number of objects in the test image is plotted along the x axis and the proportion of comparisons that result in a ``more'' response are plotted on the y-axis (blue line). Human data from 10 subjects~\cite{maldonado2020adaptation} is plotted in green. \textbf{(B)} The position of images in the embedding space fall along a straight line that starts with 0, and continues monotonically with an increasing number of objects. Thus, the position of an image in the embedding line is an estimate for the number of objects in the scene. Here we demonstrate the outputs of such a model, where we rescale the embedding coordinate (an arbitrary unit) so that one unit of distance matches the distance between the ``zero'' and the ``one'' clusters. The y-axis represents such perceived numerosity, which is not necessarily an integer value. The red line indicates perfect prediction.  Each violin plot (light blue) indicates the distribution of perceived numerosities for a given ground-truth number of objects. The width of the distributions for the higher counts indicates that perception is subject to errors. There is a slight underestimation bias for higher numbers, consistent with that seen in humans~\cite{izard2008calibrating,krueger1982single}. In fact, Krueger shows that human numerosity judgements (on images with 20 to 400 objects) follow a power function with an exponent of $0.83 \pm 0.2$. The green line and its shadow depict the range of human numerosity predictions on the same task. The orange lines are power function fits for seven models trained in the same fashion as Model B with different random initializations.
}
\label{fig:quantity_estimation_fig}
\end{figure}

Second, images are arranged almost monotonically along the embedding line according to the number of objects that are present (Fig.~\ref{fig:embedding_space}A). Thus, the representation that is developed by the model contains an order. We were curious as to whether the {\em embedding coordinate}, i.e. the position of an image along the embedding line, may be used to estimate the number of objects in the image. Any one of the features that make up the coordinates of the embedding provides a handy measure for this position, measured as the distance from the beginning of the line -- the value of these coordinates may be thought of as the firing rate of specific neurons~\cite{roitman2007monotonic}. 
We tested this hypothesis both in a relative and in an absolute quantity estimation task.  First, we used the embedding coordinate to compare the number of objects in two different images and assess which is larger, and found very good accuracy (Fig.~\ref{fig:quantity_estimation_fig}A). Second, assuming that the system may self-calibrate, e.g. by using the ``put'' action to estimate a unit of increment, then an absolute measure of quantity may be computed from the embedding coordinate.  We tested this idea by computing such a {\em perceived number} against the actual count of objects in images (Fig.~\ref{fig:quantity_estimation_fig}B). The estimates turn out to be quite accurate, with a slight underestimate that increases as the numbers become larger. Both relative and absolute estimates of quantity were accurate for as many as thirty objects (we did not test beyond this number), which far exceeds the training limit of three. We looked for image properties, other than ``number of objects'', that might drive the estimate of quantity and we could not find any convincing candidate (see Methods and Fig.~\ref{fig:covariates_fig}).

Third, image embeddings separate out into distinct ``islands'' at one end of the embedding line (Fig.~\ref{fig:embedding_space}A inset). The brain is known to spontaneously cluster perceptual information~\cite{wertheimer1938laws,harvey2013topographic}, and therefore we tested empirically whether this form of unsupervised learning may be sufficient to discover distinct categories of images/scenes from their embedding. We found that unsupervised learning successfully discovers the clusters with very few outliers in both Model A and the more challenging Model B (Fig.~\ref{fig:embedding_space}B).

Fourth, the first few clusters discovered by unsupervised learning along the embedding line are in almost perfect one-to-one correspondence with groups of images that share the same number of objects (Figs. \ref{fig:embedding_space}C). 
Once such distinct {\em number categories} are discovered, they may be used to classify images. This is because the model maps the images to the embedding, and the unsupervised clustering algorithm can classify points in the embedding into number categories. Thus, our model learns the ability to carry out instant association of images with a small set of objects with the corresponding number category. 

A fifth property of the embedding is that there is a limit to how many distinct number categories are learned. Beyond a certain number of objects one finds large clusters  which are no longer number-specific (Fig.~\ref{fig:embedding_space}). I.e. our model learns distinct categories for the numbers between zero and eight, and additional larger categories for, say, ``more than a few'' and for ``many''.

There is nothing magical in the fact that during training we limited the number of objects to three, our findings did not change significantly when we changed the number of objects that are used in training the action classifier (Fig.~\ref{fig:trainx_test8},~\ref{fig:reproducibility}), when we restricted the variability of the objects actions~(\ref{sec-jitter}), and when ``put" and ``take" could affect multiple objects at once~(\ref{sec-action_size}), i.e. when actions were imprecise. In the last two experiments, we find a small decrease in the separability of clusters in the subitization range (Figs.~\ref{fig:jitter_embedding},~\ref{fig:action_size_embedding}), such that unsupervised clustering is more sensitive to its free parameter (minimum cluster size).

\section{Discussion}
Our model and experiments demonstrate that a representation of the first few natural numbers, absolute numerosity perception, and subitization may be learned by an agent who is able to carry out simple object manipulations. The training task, action prediction, provides 
supervision for two-set comparisons. This supervision is limited to scenes with up to 3 objects, and yet the model can successfully carry out relative numerosity estimation on scenes with up to 30 objects. Furthermore, action prediction acts as a pretraining task that gives rise to a representation that can support subitization and absolute numerosity estimation without requiring further supervision.

The two mechanisms of the model, deep learning and unsupervised clustering, are computational abstractions of mechanisms that have been documented in the brain. 

A number of predictions are suggested by the regularities in the image representation that emerge from our model.

First, the model discovers the structure underlying the integers. The first few numbers, from zero to six, say, emerge as categories from spontaneous clustering of the embeddings of the corresponding images. Clustered topographic numerosity maps observed in human cortex may be viewed as confirming this prediction~\cite{harvey2013topographic}. These number categories are naturally ordered by their position on the embedding line, a fundamental property of numbers. The ability to think about numbers may be thought of as a necessary, although not sufficient, step towards counting, addition and subtraction~\cite{feigenson2004core,dehaene2009origins}. The dissociation between familiarity with the first few numbers and the ability to count has been observed in hunter-gatherer societies~\cite{pica2004exact} suggesting that these are distinct steps in cognition. In addition, we find that these properties emerge even when the number of objects involved in the action is random, further relaxing the assumptions needed for our model (Sec.~\ref{sec-action_size}).

Second, instant classification of the number of objects in the scene is enabled by the emergence of number categories in the embedding, but it is restricted to the first few integers. This predicts a well-known capability of humans,  commonly called {\em subitization}~\cite{jevons1871power,burr2011adaptation}.  

Third, a linear structure, which we call {\em embedding line}, where images are ordered according to quantity, is an emergent representation. This prediction is strongly reminiscent of the {\em mental number line} which has been postulated in the psychology literature~\cite{restle1970speed,dehaene1993mental,dehaene2004arithmetic,rugani2015number}. The embedding line confers to the model the ability to estimate quantities both in relative comparisons and in absolute judgments. The model predicts the ability to carry out relative estimation, absolute estimation, as well as the tendency to slight underestimation in absolute judgments. These predictions are confirmed in the psychophysics literature~\cite{burr2008visual,izard2008calibrating}. 

Fourth, subitization and numerosity estimation extend far beyond the number of objects used in training. While the model trains itself to classify actions using up to three objects, subitization extends to 5-8 objects and numerosity estimation extends to at least thirty, which is as far as we tested. Extrapolating from the training set is a hallmark of abstraction, which eludes most supervised models~\cite{trask2018neural}, yet has been shown in rhesus monkeys~\cite{cantlon2006shared}. Consensus in the deep networks literature is that models {\em interpolate} their training set, while here we have a striking example of generalization {\em beyond} the training set.

Fifth, since in our model manipulation teaches perception, one would predict that children who lack the ability or the drive to manipulate would show retardation in the development of a number sense. A study of children with Developmental Coordination Disorder~\cite{gomez2015mathematical} is consistent with this prediction.

Sixth, our model predicts that adaptation affects estimation, but not subitzation. This is because subitization solely relies on classifiers, which allows for a direct estimate of quantity. Estimation, however, relies on an analog variable, the coordinate along the embedding line, which requires calibration. These predictions are confirmed in the psychophysics literature~\cite{izard2008calibrating,burr2008visual}. 

Seventh, our model predicts the existence of summation units, which have been documented in the physiology literature~\cite{roitman2007monotonic} and have been postulated in previous models~\cite{verguts2004representation}. It does not rule out the simultaneous presence of other codes, such as population codes or labeled-line codes~\cite{nieder2016neuronal}. 

The model is simple and our clustering method is essentially {\em parameter-free}. Our observations are robust with respect to large variations in the dimension of the embedding, the number of objects in the training set and the tuning parameters of the clustering algorithm. Yet, the model accounts qualitatively and, to some extent, quantitatively for a disparate set of observations by psychologists, psychophysicists and cognitive scientists.

There is a debate in the literature on whether estimation and subitization are supported by the same mechanisms or separate ones~\cite{burr2008visual,cheyette2020unified}. Our model suggests a solution that supports both arguments: both perceptions rely on a common representation, the embedding. However, the two depend on different mechanisms that take input from this common representation. 

It is important to recognize the limitations of our model: it is designed to explore the minimal conditions that are required to learn several cognitive number tasks, and abstracts over the details of a specific implementation in the brain.
For instance, we limit the model to vision, while it is known that multiple sensory systems may contribute, including hearing, touch and self-produced actions~\cite{amalric2018role,crollen2020visual,anobile2020sensorimotor}. Furthermore, the visual system serves multiple tasks, such as face processing, object recognition, and navigation. Thus, it is likely that multiple visual maps are simultaneously learned, and it is possible that our ``latent representation'' is shared with other visual modalities~\cite{2019numbnasrer}.  Additionally, we postulate that visually-guided manipulation, and hence the ability to detect and locate objects, is learned before numbers. Thus, it would perhaps be more realistic to consider input from an intermediate map where objects have been already detected and located, and are thus represented as ``tokens'', in visual space, and this would likely make the model's task easier, perhaps closer to Model A than to Model B. However, making this additional assumption is not necessary for our observations.

An interesting question is whether object manipulation, which in our model acts as the supervisory signal during play, may be learned without supervision and before the learner is able to recognize numbers.  Our work sheds no light on this question, and simply postulates that this signal is available and, importantly, that the agent is able to discriminate between the three put, take and shake actions. Our model shows that this simple signal on scenes containing a few objects may be bootstrapped to learn about integers, and to perform subitization and numerosity estimation in scenes containing many objects.

Our investigation adds a concrete case study to the discussion on how abstraction may be learned without explicit supervision. While images containing, say, five objects will look very different from each other, our model discovers a common property, i.e. the number of items, which is not immediately available from the brightness distribution or other scene properties.  The mechanism driving such abstraction may be interpreted as an implicit {\em contrastive learning} signal~\cite{hadsell2006dimensionality}, where the {\em shake} action identifies pairs of images that ought to be considered as similar, while the {\em put} and {\em take} actions signal pairs of images that ought to be considered dissimilar, hence the clustering. However, there is a crucial difference between our model and traditional contrastive learning. In contrastive learning, the similarity and dissimilarity training signals are pre-defined for each image pair and the loss is designed to achieve an intended learning goal -- to bring the embeddings of similar images together and push the embeddings of dissimilar images apart. In our model, image pairs are associated by an action and the network is free to organize the embeddings in any manner that would be efficient for solving the action prediction task. The learned representation is surprisingly robust -- while the primary supervised task, action classification, does not generalize well beyond the three objects used in training, the abstractions of number and quantity extend far beyond it.

\section{Methods}

\subsection{Network Details}
The network we train is a standard deep network~\cite{lecun2015deep} composed of two stages. First, a feature extraction network maps the original image of the scene into an embedding space (Fig. \ref{fig:network}A). Second, a classification network takes the embedding of two sequential images and predicts the action that modified the first into the second (Fig. \ref{fig:network}B). Given the fact that the classification network takes the embedding of two distinct images as its input, each computed by identical copies of the feature extraction network, the latter is trained in a Siamese configuration~\cite{bromley1994signature}.

The feature extraction network is a 9-layer CNN followed by two fully connected layers (details in Fig.~\ref{fig:network_details}A). The first 3 layers of the feature extraction network are from AlexNet~\cite{krizhevsky2012imagenet} pre-trained on ImageNet~\cite{deng2009imagenet} and are not updated during training. The remaining four convolutional layers and two fully connected layers are trained in our action prediction task.

The dimension of the output of the final layer is a free parameter (it corresponds to the number of features and to the dimension of the embedding space). In a control experiment we varied this dimension from one to 256, and found little difference in the action classification error rates (Fig.~\ref{fig:embedding_dimension_fig}). We settled for a two-dimensional output for the experiments that are reported here.

The classification network is a two-layer fully connected network that outputs a three-dimensional one-hot-encoding vector indicating a put, take or shake action (details in Fig.~\ref{fig:network_details}B). 

\subsubsection{Training procedure}
The network was trained with a negative log-likelihood loss (NLL loss) function with a learning rate of 1e-4. The NLL loss calculates error as the -log of the probability of the correct class. Thus, if the probability of the correct class is low (near 0), the error is higher. The network was trained for 30 epochs with 30 mini-batches in each epoch. Each mini-batch was created from a sequence of 180 actions, resulting in 180 image pairs. Thus, the network saw a total of 162,000 unique pairs of images over the course of training.

We tested for reproducibility by training Model B thirty times with different random initializations of the network and different random seeds in our dataset generation algorithm. The embeddings for these reproduced models are shown in Figure~\ref{fig:reproducibility}. 

\subsubsection{Compute}
All models were trained on a GeForce GTX TITAN X using PyTorch. Each model takes at most 20 minutes to train. We train a total of 106 models (including supplemental experiments).

\subsection{Synthetic Dataset Details}
\subsubsection{Training sets}
\label{sec:training-stats}
We carried out experiments using synthetic image sequences where objects were represented by randomly positioned squares. The images were 244x244 pixels (px) in size. Objects were positioned with uniform probability in the image, with the exception that they were not allowed to overlap and a margin of at least 3px clearance between them was imposed. We used two different statistics of object appearance: identical size (15px) and contrast (100\%) in the first, and variable size (10px - 30px) and contrast (9.8\% - 100\%) in the second (Fig.~\ref{fig:main_data}). Mean image intensity statistics for the two training sets are shown in Figure \ref{fig:dataset_statistics}. The mean image intensity is highly correlated with the number of objects in the first dataset, while it is ambiguous and thus not very informative in the second. We elaborate on covariates like mean image intensity in the following section.

Each training sequence was generated starting from zero objects, and then selecting a random action (put, take, shake) to generate the next image. The take action is meaningless when the scene contains zero objects and was thus not used there. We also discarded put actions when the objects reached a maximum number. This limit was three for most experiments, but limits of five and eight objects were also explored (Fig.~\ref{fig:trainx_test8}).

\subsubsection{Test sets}
In different experiments we allowed up to eight objects per image (Figs. \ref{fig:errors}, \ref{fig:trainx_test8}) and thirty objects per image (Figs.~\ref{fig:embedding_space},~\ref{fig:quantity_estimation_fig}A,~\ref{fig:quantity_estimation_fig}B) in order to assess whether the network can generalize to tasks on scenes containing previously unseen numbers of objects. The first test set (up to 8 objects) was generated following the same recipe as the training set. The second test (up to 30 objects) set was generated to have random images with the specified number of objects (without using actions), this test set is guaranteed to be balanced. In section~\ref{sec-covariates}, we use the 30 object test set to estimate covariates for numerosity and analyze their impact on task performance. We were unable to find an image property that would ``explain away'' the abstraction of number (Fig. \ref{fig:covariates_fig}). We note that a principled analysis of the information that is carried out by individual object images is still missing from the literature~\cite{testolin2020visual} and this point deserves more attention. 

\subsection{Action classification performance}
To visualize how well the model was able to perform the action classification task, we predict actions between pairs of images in our first test set. The error, calculated by comparing the ground truth actions to the predicted actions, is plotted with respect to the number of objects in the visual scene at $x_t$. 95\% Bayesian confidence intervals with a uniform prior were computed for each data point, and a lower bound on the number of samples is provided in the figure captions (Figs. \ref{fig:errors}, \ref{fig:embedding_dimension_fig}, \ref{fig:trainx_test8}). 

\subsection{Interpreting the embedding space}
We first explored the structure of the embedding space by visualizing the image embeddings in two dimensions.  The points, each one of which corresponds to one image, are not scattered across the embedding. Rather, they are organized into a structure that exhibits five salient features: (a) the images are arranged along a one-dimensional structure, (b) the ordering of the points along the line is (almost) monotonic with respect to the number of objects in the corresponding images, (c) images are separated into groups at one end of the embedding, and these groups are discovered by unsupervised learning, (d) these first few clusters are in one-to-one correspondence with the first few natural numbers, (e) there is a limit to how many number-specific clusters are discovered (Fig. \ref{fig:embedding_space}).

To verify that the clusters can be recovered by unsupervised learning we applied a standard clustering algorithm, and found almost perfect correspondence between the clusters and the first few natural numbers (Fig.~\ref{fig:embedding_space}). The clustering algorithm used was the default Python implementation of HDBSCAN \cite{McInnes2017}. HDBSCAN is a hierarchical, density based clustering algorithm, and we used the euclidean distance as an underlying metric \cite{campello2013density}. HDBSCAN has one main free parameter, the minimum cluster size, which was set to 90 in Figure \ref{fig:embedding_space}. All other free parameters were left at their default values. Varying the minimum cluster size between 5 and 95 does not have an effect on the first few clusters, although it does create variation in the number and size of the later clusters. Beyond 95, the algorithm finds only three clusters corresponding to 0, 1 and greater than 1.

One additional structure is not evident from the the embedding and may be recovered from the action classifier: the connections between pairs of clusters. For any pair of images that are related by a manipulation, two computations will be simultaneously carried out; first, the supervised action classifier in the model will classify the action as either P, T, or S (Fig.~\ref{fig:errors}) and, at the same time, the unsupervised subitization classifier (Fig.~\ref{fig:embedding_topology_histogram}A) will assign each image in the pair to the corresponding number-specific cluster. As a result, each pair of images that is related by a P action provides a directed link between a pair of clusters (Fig.~\ref{fig:embedding_topology_histogram}A, red arrows), and following such links one may traverse the sequence of numbers in an ascending order. The T actions provide the same ordering in reverse (blue arrows). Thus, the clusters corresponding to the first few natural numbers are strung together like the beads in a necklace, providing an unambiguous ordering that starts from zero and proceeds through one, two etc. (Fig.~\ref{fig:embedding_topology_histogram} A, B). The numbers may be visited both in ascending and descending order. As we pointed out earlier, the same organization may be be obtained more simply by recognizing that the clusters are spontaneously arranged along a line, which also supports the natural ordering of the numbers~\cite{dehaene1995towards,zorzi2002neglect,dehaene2004arithmetic}. However, the connection between the order of the number concepts, and the actions of put and take, will support counting, sum and subtraction.

To estimate whether the embedding structure is approximately one dimensional and linear in higher dimensions we computed the one-dimensional linear approximation to the embedding line, and measured the average distortion of using such approximation for representing the points. More in detail, we first defined a mean-centered embedding matrix with M points and N dimensions, each point corresponding to the embedding of an image. We then computed the best rank 1 approximation to the data matrix by computing its singular value decomposition (SVD) and zeroing all the singular values beyond the first one. If the embedding is near linear, this rank 1 approximation should be quite similar to the original matrix. To quantify the difference between the original matrix and the approximation, we calculated the element-wise residual (the Frobenius norm of the difference between the original matrix and the approximation), then computed the ratio of the Frobenius norm of the residual matrix and the Frobenius norm of the original matrix. The nearer the ratio is to 0, the smaller the residual, and the better the rank 1 approximation. We call this ratio the {\em linear approximation error}, we show this error compared to some embeddings in Figure~\ref{fig:reproducibility}. We computed the embedding for dimensions 8, 16, 64, and 256, (one experiment each) and found ratios of 0.702\%, 2.23\%, 2.77\%, and 2.24\%, suggesting that they are close to linear. 

\subsection{Estimating relative quantity}
We can use the perceived numerosity to reproduce a common task performed in human psychophysics. Subjects are asked to compare a reference image to a test image and respond in a two-alternative forced choice paradigm with ``more'' or ``less''. We perform the same task using the magnitude of the embedding as the fiducial signal. The model responds with more if the embedding of the test image has a larger perceived numerosity than the reference image. The psychometric curves generated by our model are presented in Figure \ref{fig:quantity_estimation_fig}A and match qualitatively the available psychophysics~\cite{burr2008visual, krueger1982single}.

\subsection{Estimating absolute quantity}
As described above, the clusters are spaced regularly along a line and the points in the embedding are ordered by the number of objects in the corresponding images (Fig. \ref{fig:embedding_topology_histogram}). We postulate that the number of objects in an image is proportional to the distance of that image's embedding from the embedding of the empty image. Given the linear structure, any one of the embedding features, or their sum, may be used to estimate the position along the embedding line. In order to produce an estimate we use the embedding of the ``zero" cluster as the origin. The zero cluster is special, and may be detected as such without supervision, because all its images are identical and thus it collapses to a point. The distance between ``zero" and ``one", computed as the pairwise distance between points belonging to the corresponding clusters, provides a natural yardstick. This value, also learned without further supervision, can be used as a unit distance to to interpret the signal between 0 and n. This estimate of numerosity is shown in Figure \ref{fig:quantity_estimation_fig}B against the actual number of objects in the image. We draw two conclusions from this plot. First, our unsupervised model allows an estimate of numerosity that is quite accurate, within 10-15\% of the actual number of objects. Second, the model produces a systematic underestimate, similar to what is observed psychophysically in human subjects~\cite{izard2008calibrating}. 

\section{Dataset \& Code Availability}
All data generated or analysed during this study can be found \href{https://data.caltech.edu/records/achmg-dc274}{here}. The data can also be generated with the code. Code is available \href{https://github.com/nkondapa/ManipulationToNumberSense}{here}.
 
\section{Acknowledgements}
The California Institute of Technology and the Simons Foundation (Global Brain grant 543025 to PP) generously supported this work. Daniel Israel wrote the code for the jitter and action size supplemental experiments. We are very grateful to a number of colleagues who  provided references to the literature and insightful suggestions: Alessandro Achille, Katie Bouman, David Burr, Eli Cole, Jay MacClelland, Markus Meister,  Mario Perona, Giovanni Paolini,  Stefano Soatto, Alberto Testolin, Kate Stevenson, Doris Tsao, Yisong Yue and two anonymous referees.

\bibliographystyle{unsrt}
\bibliography{references}
\nocite{deng2009imagenet, krizhevsky2012imagenet}

\include{supp}

\end{document}

%% file: supp.tex




\clearpage
\newpage
\renewcommand{\thesection}{\Alph{section}}
\renewcommand{\thesuppfigure}{S\arabic{suppfigure}}

\setcounter{section}{0} 




\section{Additional Experiments}

\subsection{Controlling for spurious correlates of ``number''}
\label{sec-covariates}
Do image properties, other than the abstraction of ``object number'', drive the quantity estimate of our model? Many potential {\em confound variables}, such as the count of pixels that are not black, are correlated with object number and might play a role in the model's ability to estimate the number of objects in the scene. If that were the case, one might argue that our model is not learning the abstraction of ``number'', but rather learning to measure image properties that are correlated with number.

We controlled for this hypothesis by exploiting the natural variability of our test set images. We explored three image properties that correlate with the number of objects and might thus be exploited to estimate the number of objects: (a) overall image brightness, (b) the area of the envelope of the objects in the image, and (c) the total number of pixels that differ from the background. Since objects in training set B vary both in size and in contrast, these three variables are not deterministically related to object number and thus, we reason, counfound variable fluctuations ought to affect error rates independently of the number of objects.

We focused on close-call relative estimate tasks (e.g. 16 vs 18 objects), where errors are frequent both for our model and for human subjects, and, while holding the number of objects constant in each of the two scenes being compared, we studied the behavior of error rates as a function of fluctuations in the confound variables. One would expect more errors when comparing image pairs where quantities that typically correlate with the number of objects are anticorrelated in the specific example (Fig.~\ref{fig:covariates_fig_examples}). Conversely, one would expect lower error rates when the confound variables are positively correlated with number.

In Fig.~\ref{fig:covariates_fig} error rates are plotted vs each one of the confound variables when the n. of objects is held constant. We could not find large systematic biases even for extreme variations in the confound variables. In conclusion, we do not find support for the argument that any of the confound variables we studied is implicated significantly in the estimate of quantity.

\begin{suppfigure}
\begin{center}
    \renewcommand{\thesubsuppfigure}{\Alph{subsuppfigure}}
    \includegraphics[width=1\linewidth, trim={0 1.6cm 0 0}, clip]{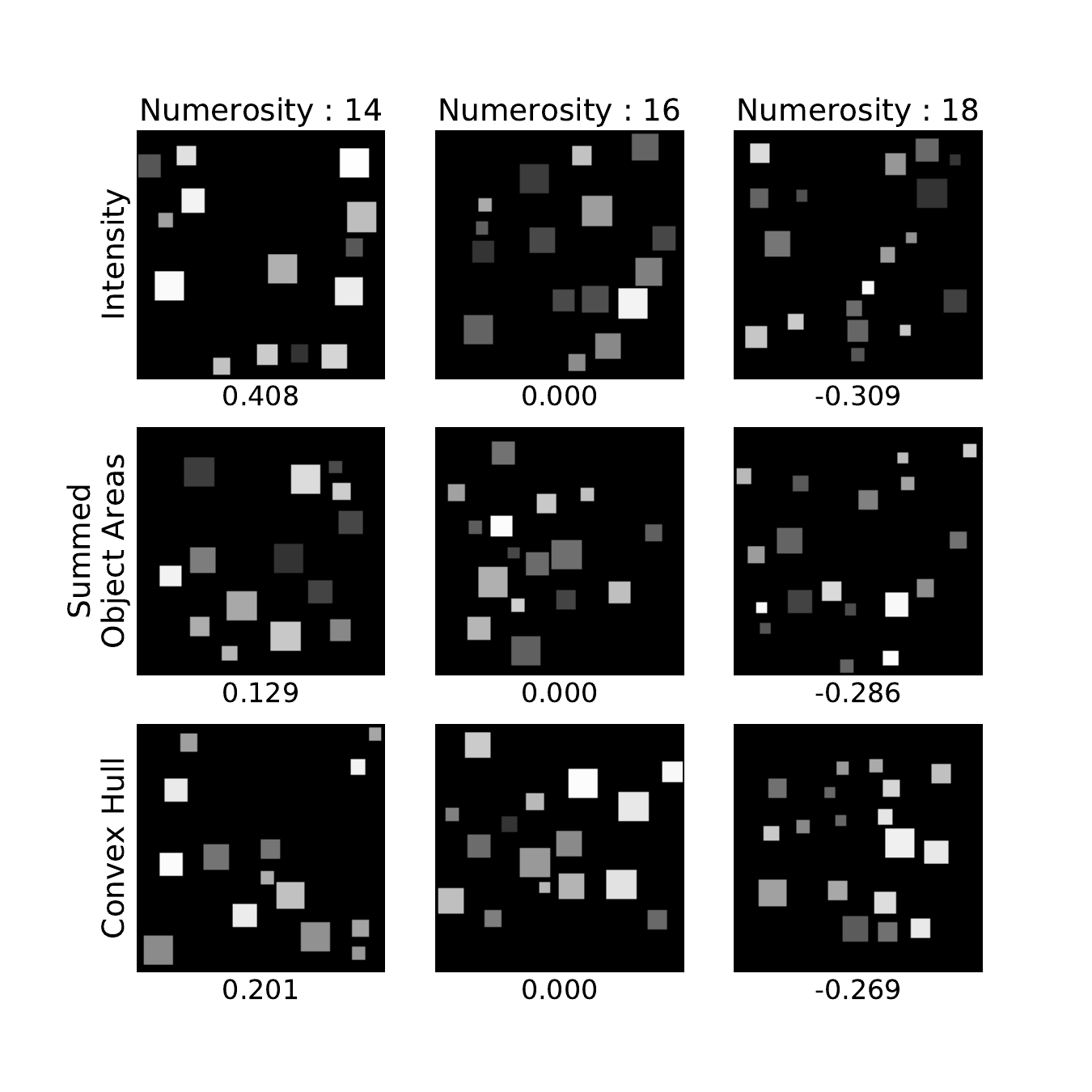}
\end{center}
\linespread{1} \caption{\textbf{Sample images where covariates are anticorrelated with number.} We sample images where the three covariates we study (one covariate per row) are anticorrelated with the number of objects. The number below each plot shows the fractional difference from the value of the covariate in the reference image (center column). For example, in the top right, there is a 30.9\% decrease in average image intensity when compared to the intensity in the reference image (center column). Another example: in the last row, the scene with 18 objects has a 26.9\%  smaller convex hull than the corresponding scenes with 14 and 16 objects. For each row, from the lowest numerosity to the highest, the model predicts a perceived numerosity of 12.82, 14.01, and 16.60 (Intensity); 13.21, 14.43, 15.55 (Summed Object Area); 13.22, 15.28, 16.44 (Convex Hull). Thus, our model correctly classifies the relative numerosity for each one of the image pairs that may be formed from each row (our model slightly underestimates numerosity, see Figure \ref{fig:quantity_estimation_fig}B.) Image pairs formed this way are used in the experiments shown in Figure \ref{fig:covariates_fig}, where this manipulation was repeated multiple times and confidence intervals were computed. } 
\label{fig:covariates_fig_examples}
\end{suppfigure}
\begin{suppfigure}
\begin{center}
    \renewcommand{\thesubsuppfigure}{\Alph{subsuppfigure}}
    \includegraphics[width=1\linewidth]{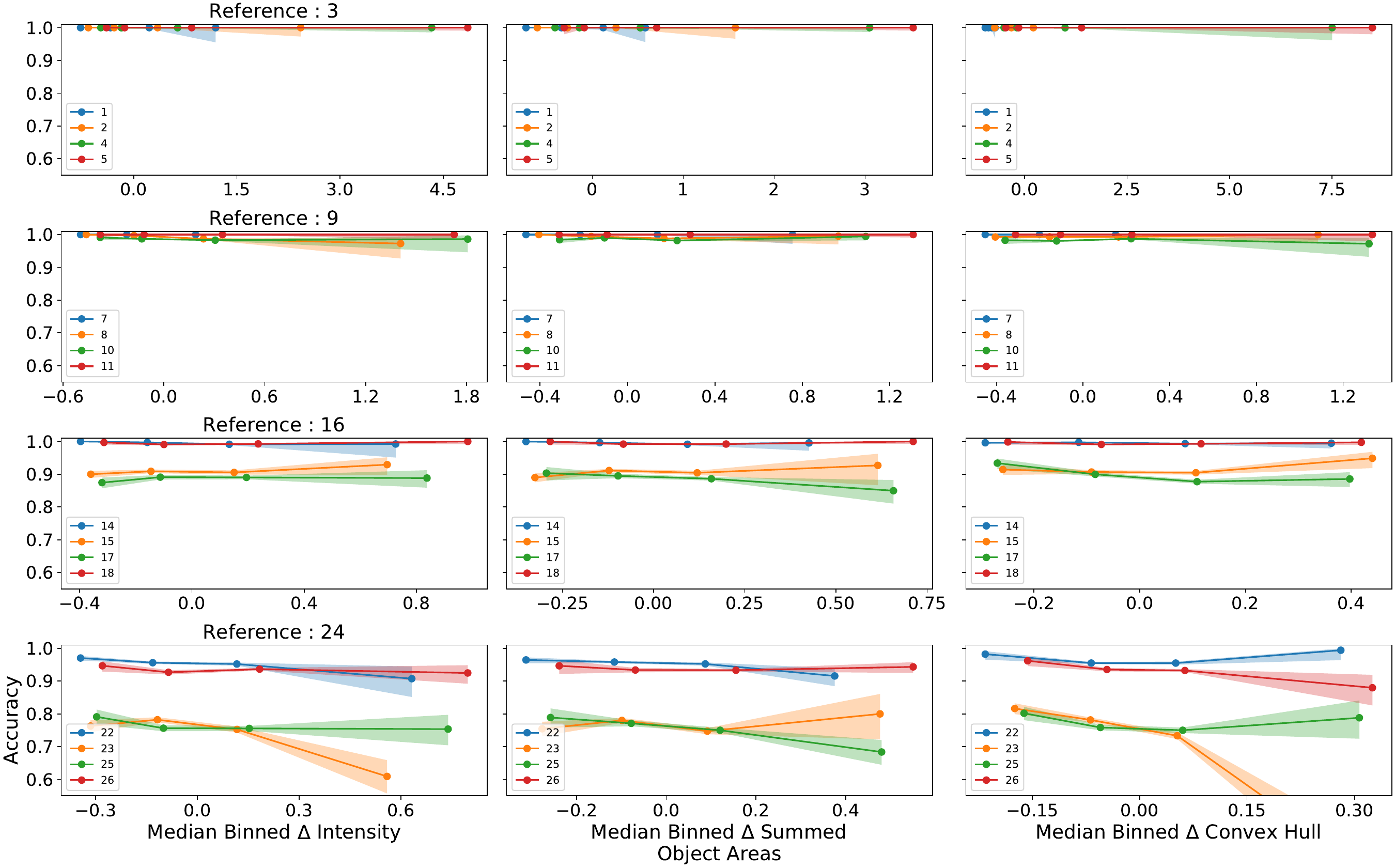}
\end{center}
\linespread{1} \caption{\textbf{Effects of covariates of numerosity.} Three covariates of the number of objects in the scene are explored for possible influence on our model's estimate of numerosity. These are average image intensity {\bf (left column)}, the sum of the areas of the objects {\bf (middle column)}, and the area of the objects' convex hull {\bf (right column)}. Each plot shows the error rates in a relative quantity discrimination task like the one in Figure \ref{fig:quantity_estimation_fig}A. We generate a test set of 4650 test images, 150 images per number of objects. For each plot we chose reference images containing respectively 3, 9, 16 and 24 objects (rows of the figure) and had our model judge relative numerosity w.r. to test images containing a different but similar number of objects (indicated in the legend and associated with colors). Given the stochastic nature of the images, the covariates vary over a wide range for each number of objects (see examples in Fig. \ref{fig:covariates_fig_examples}). For each number of objects, we plot the model's error rates (y axis) as a function of the value of the covariate quantity (x axis) which is expressed as fractional difference from the reference image (the values are binned). Shadows display 95\% Bayesian confidence intervals($N > 100$, where N is bin size). Horizontal error lines indicate no correlation of numerosity estimation with the covariate quantity.  A few lines have slopes that differ slightly from zero indicating a possible correlation. However, some of the slopes indicate a negative correlation (i.e. the better the signal, the higher the error rate). From this evidence it is difficult to conclude that that the model is exploiting anything but ``number'' to estimate numerosity.
} 
\label{fig:covariates_fig}
\end{suppfigure}

\subsection{Interpreting the Embedding Space}

\begin{suppfigure}
\begin{center}
    \includegraphics[width=0.8\linewidth]{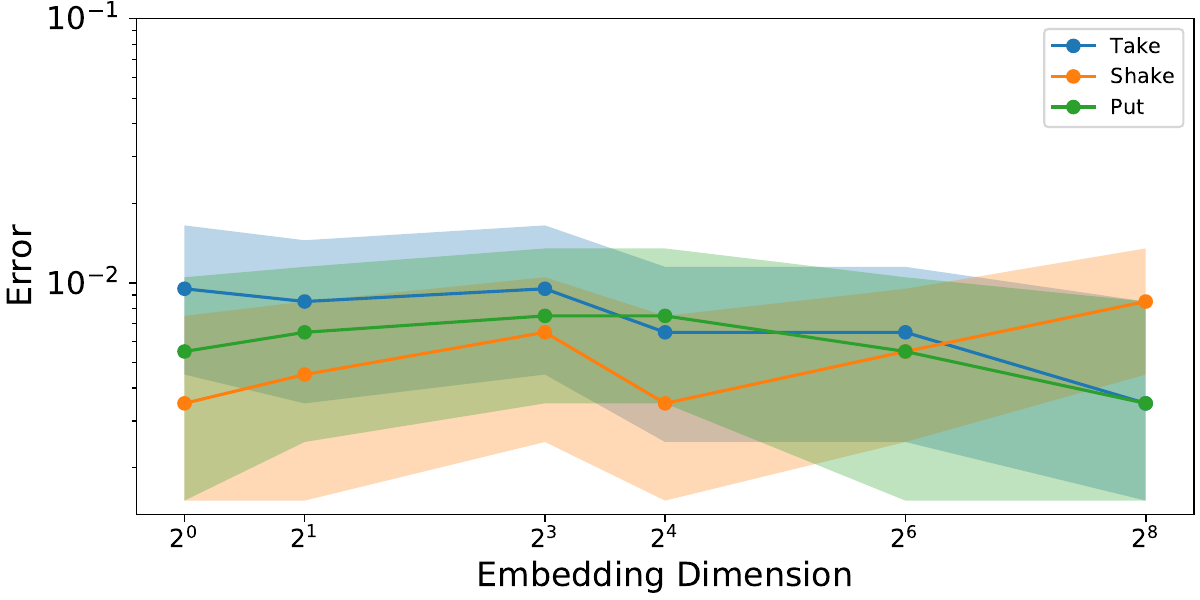}
\end{center}
\linespread{1} \caption{\textbf{Action classification error as a function of embedding dimension.} Classification errors for Model B, averaged over the number of items in the scene (0 - 3) are plotted as a function of the dimension of the embedding (a free parameter in our model). Since the effect is minimal we arbitrarily picked a dimension of two for ease of visualization (Figs.~\ref{fig:embedding_space},~\ref{fig:embedding_topology_histogram}). The shadows show 95\% Bayesian confidence intervals (287 $\leq$ N $\leq$ 355).}
\label{fig:embedding_dimension_fig}
\end{suppfigure}

Does the dimension of the embedding space influence the action classification error? We wondered what is the effect of this free parameter on the model's performance.  We explored this question by training our model repeatedly with the same training images, and varying the dimension of the embedding (Fig.~\ref{fig:network}). Figure~\ref{fig:embedding_dimension_fig} shows that the effect of the embedding dimension is negligible. This was initially surprising to us. An explanation may be found in the fact that learning produces an embedding that is organized as a line (see Fig~\ref{fig:embedding_space} and Sec.~\ref{sec:reproducibility}).

\begin{suppfigure}
\begin{center}
    \includegraphics[width=1\linewidth]{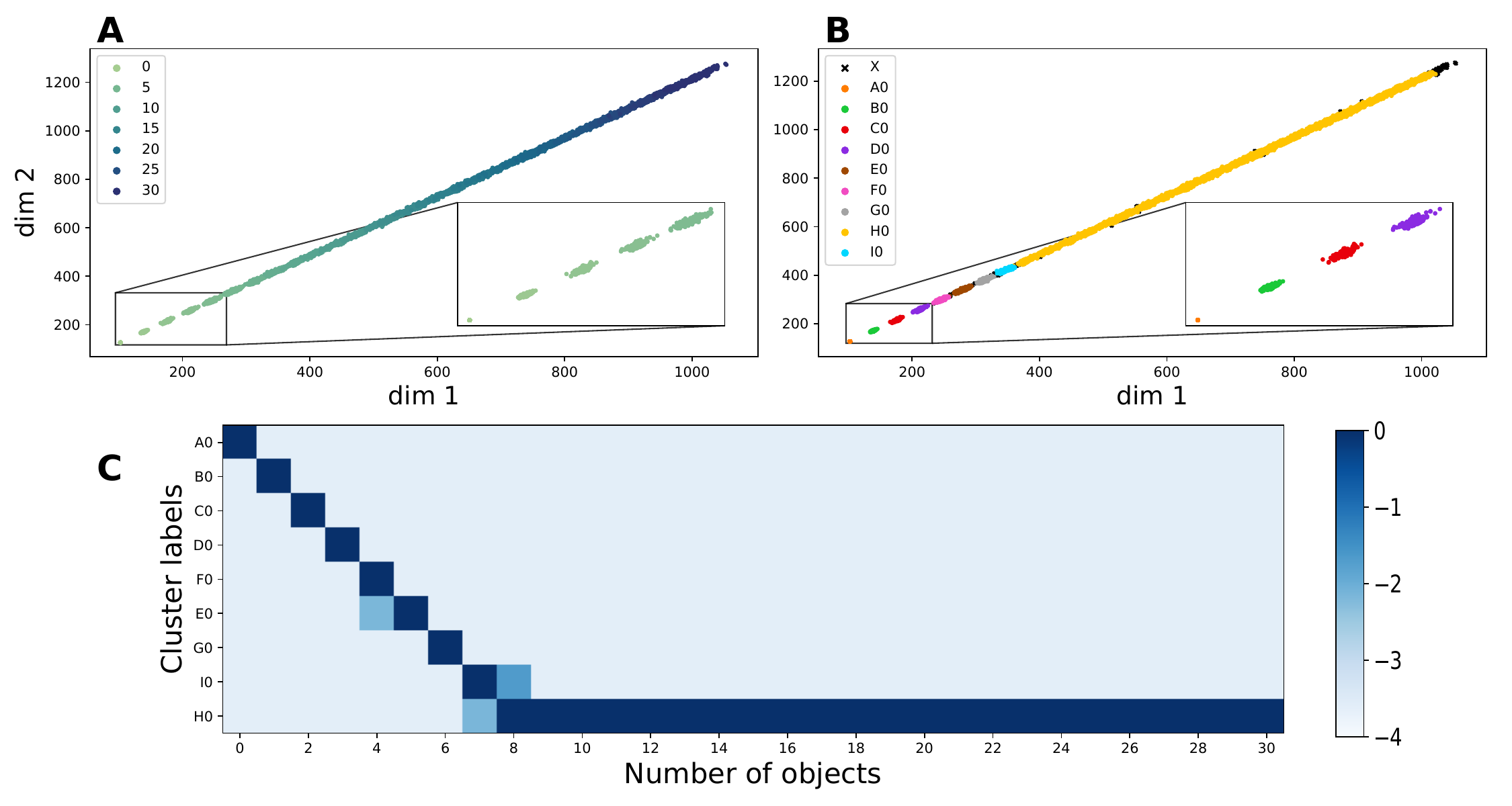}
    \renewcommand{\thesubsuppfigure}{\Alph{subsuppfigure}}
\end{center}
\linespread{1} \caption{\textbf{The embedding space for Model A.} We reproduce Fig.~\ref{fig:embedding_space} for model A. \textbf{(A)} Similar to Model B, we observe a monotonically increasing line with well seperated groups at lower quantities. \textbf{(B)} We apply an unsupervised clustering algorithm to the embeddings. Each cluster that is discovered is denoted by a specific color. The cluster X, denoted by black crosses, indicates points that the clustering algorithm excluded as outliers. 
\textbf{(C)} The confusion matrix shows that the clusters that are found by the clustering algorithm correspond to numbers. Images containing 0 - 7 objects are neatly separated into individual clusters; after that images are collected into a large group that is not in one-to-one correspondence with the number of objects in the image. The color scale is logarithmic (base 10).} 
\label{fig:embedding_space_a}
\end{suppfigure}

\begin{suppfigure}
\begin{center}
    \renewcommand{\thesubsuppfigure}{\Alph{subsuppfigure}}
    \includegraphics[width=0.75\linewidth]{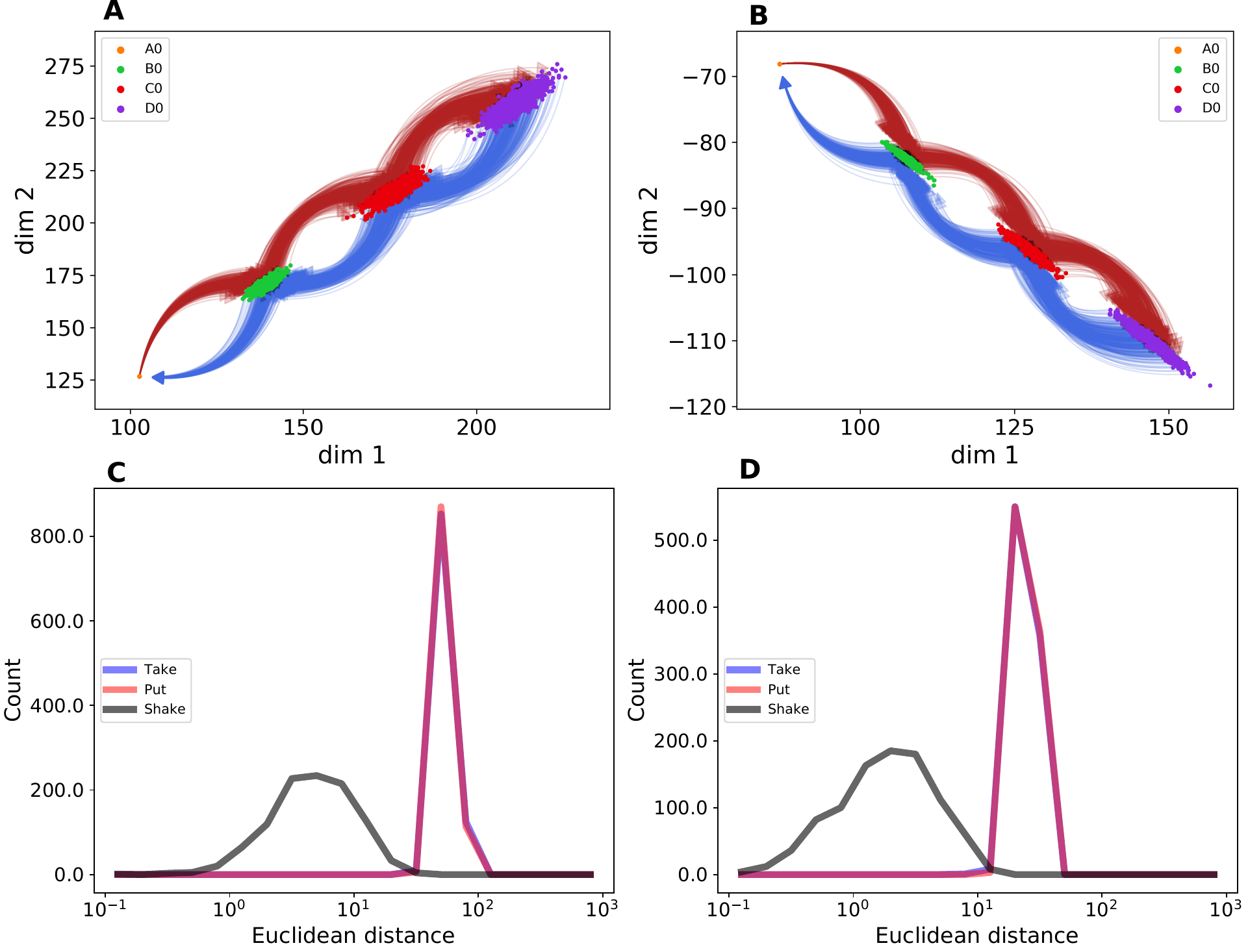}
\end{center}
\linespread{1} \caption{\small \textbf{Embeddings with topology for Model A and Model B.} A close-up look at the embedding space within the training limit. The left side are plots from Model A and the right side from  Model B. \textbf{(A), (B)} Unsupervised clustering is performed on the embedding space. Each embedding is colored by it's cluster. Each cluster A0 - D0 correspond to images with numerosities 0 - 3. The clusters are well-separated. The ``zero" clusters, for both Model A and Model B, are immediately recognizable as they have no variance (orange dot). As numerosity increases, Model A clusters remain  well-separated, whereas Model B clusters begin to come closer to each other. We also overlay a topology from the training actions (P), (T), (S). Blue arrows joining a pair of points represent take actions, red arrows represent put actions. Arrows representing shake actions are under the point clouds and are mostly not visible. \textbf{(C), (D)} Distances between pairs of points in the embedding space are histogrammed by action. The histograms show the clearly different distribution for shake actions in comparison to take and put actions. Furthermore, the overlap between shake and non-shake actions is smaller for Model A than Model B, explaining the higher performance in action classification for Model A.}
\label{fig:embedding_topology_histogram}
\end{suppfigure}


Next, we explored the structure of the embedding space in the region where images containing 0-3 objects (the training range) are represented. As discussed in the main text we find that the embedding is organized into clusters (Fig.~\ref{fig:embedding_topology_histogram} (A,B)). Each cluster contains embeddings of images with the same number of objects. For each pair of images that were generated by a {\em put} action we drew a red arrow connecting the corresponding embeddings. We used blue arrows for {\em take} pairs. It is clear from the figure that by following the red arrows one may visit numbers in increasing order: 0-1-2-3 and vice-versa for blue arrows, i.e. the embedding that is produced by our model supports counting up and down.

\subsection{Varying Training Limit}

\begin{suppfigure}[h]
\begin{center}
    \renewcommand{\thesubsuppfigure}{\Alph{subsuppfigure}}
    \begin{subsuppfigure}{0.95\textwidth}
        \hspace*{-0.5cm} 
        \includegraphics[width=1\linewidth]{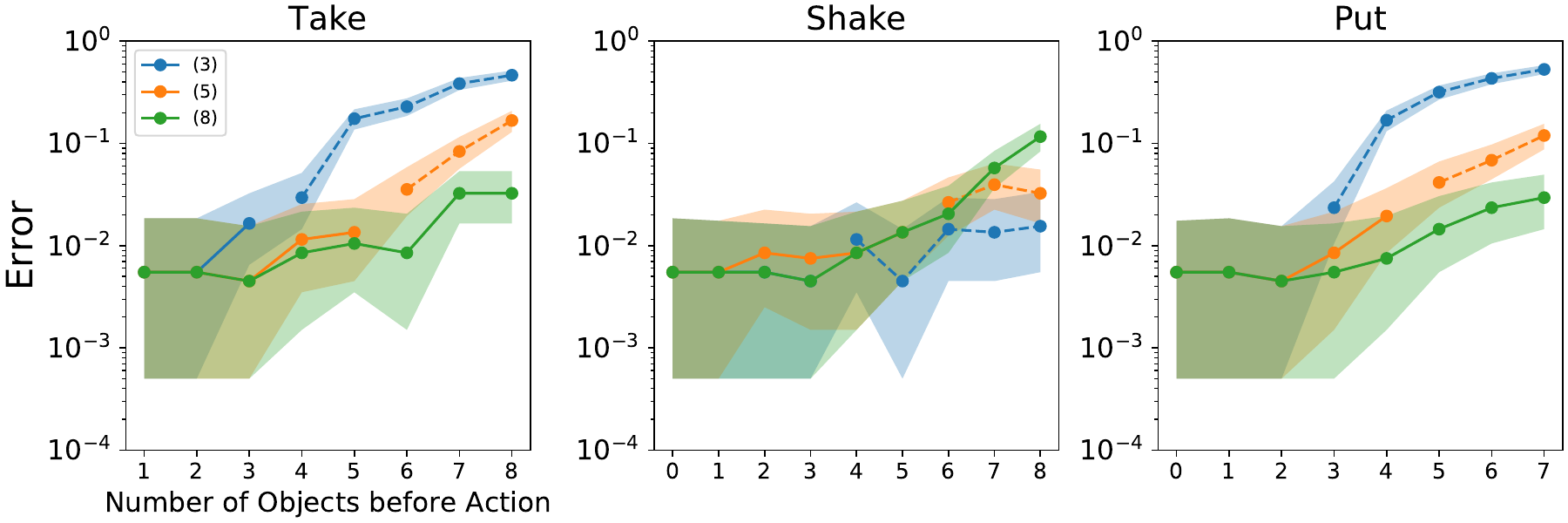}
    \end{subsuppfigure}
\end{center}
\linespread{1} \caption{\textbf{Effect of modifying the training limit.} (see also Fig.~\ref{fig:errors}) In order to explore the effect of the number of objects during training, we trained the network to predict actions using a maximum of 3, 5, or 8 objects with images like those in dataset B (Fig. \ref{fig:main_data}B). We tested the network on 8 objects. Each panels shows errors on the training task and are in the same style as Figure \ref{fig:errors}. The line-breaks and dashed lines mark where the training limit ends and the testing region begins, and the legend shows the training limit in parentheses. The shadows provide 95\% confidence intervals (287 $\leq$ N $\leq$ 355). As expected, the error is lower when the training limit is higher. 
}
\label{fig:trainx_test8}
\end{suppfigure}

In our main experiment we trained our model to classify actions with scenes containing from zero to three objects. Does this choice influence qualitatively or quantitatively our observations? 

To explore this question we re-trained our model using images that were generated with a total number of three, five and eight objects. As expected, we find that adding more objects to the training images reduces the action classification error for image pairs with corresponding number of objects (Fig.~\ref{fig:trainx_test8}). We find no change in the linearity of the embeddings, however, the number of clusters seems to increase with the training limit (Figs.~\ref{fig:reproducibility}A,B). This increase in clusters that corresponds with training limit likely explains the improvement in action classification performance. 

\subsection{Reproducibility of the 1D structure of the embedding}
\label{sec:reproducibility}

\begin{suppfigure}[h]
\begin{center}
    \renewcommand{\thesubsuppfigure}{\Alph{subsuppfigure}}
    \begin{subsuppfigure}{1\textwidth} \includegraphics[width=1\textwidth, trim={0 2.2cm 0 0}, clip]{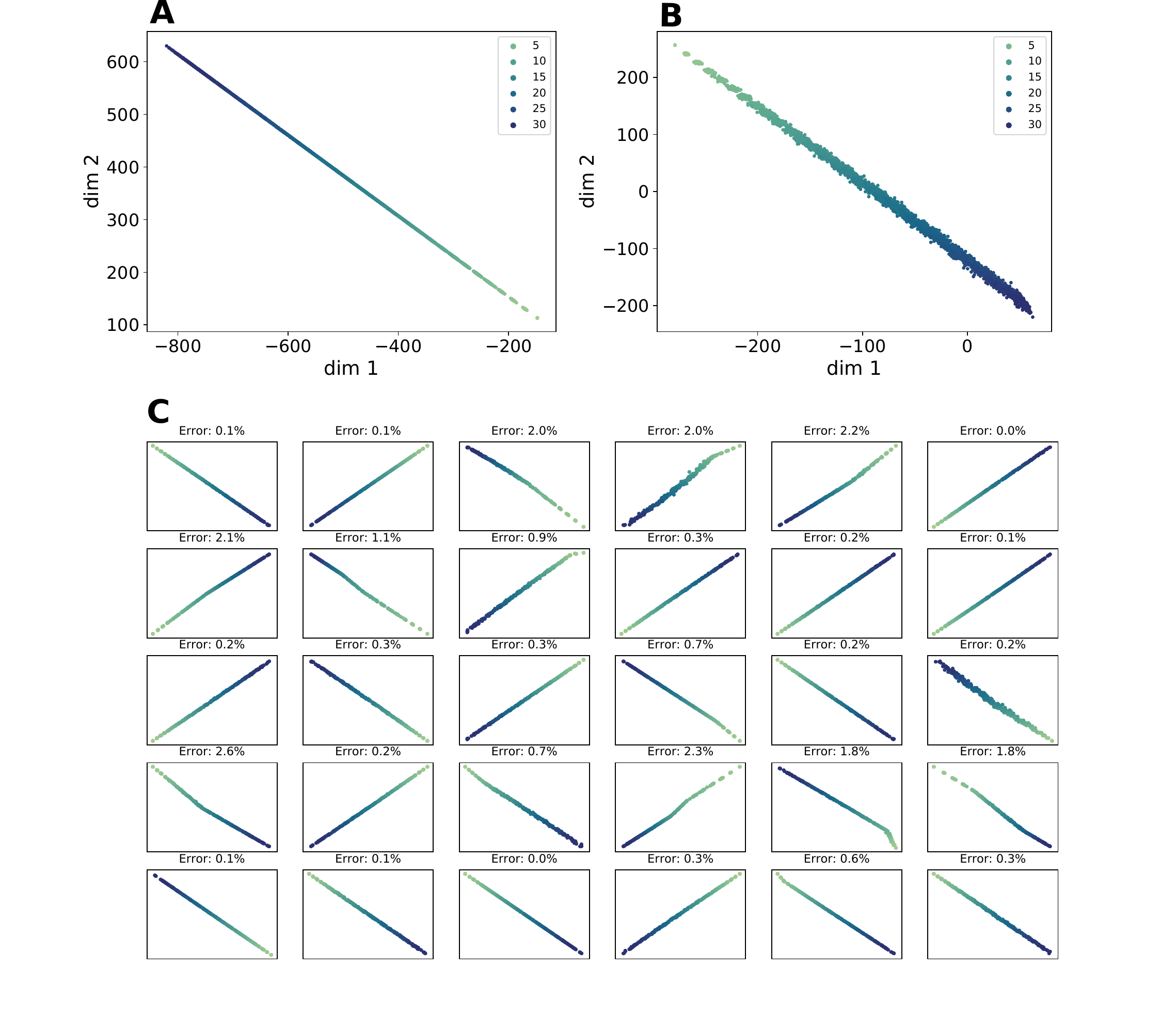}
    \end{subsuppfigure}
\end{center}
\linespread{1} \caption{\textbf{Miscellaneous embedding spaces.} (see also Fig.~\ref{fig:embedding_space}) \textbf{(A)}  Embedding space for the network trained on dataset B,  with up to five objects. \textbf{(B)} Embedding space for the network trained on dataset B, with up to eight objects.
\textbf{(C)} Embedding spaces for 30 different random initializations. We repeated the training procedure 30 times on different random initializations of dataset B, with a training limit of 3 objects. Qualitatively, 21 embedding spaces look like a straight line, six initializations present a slight kink in the line, and three instances either present a large kink or two kinks. The linear approximation error (Methods - Interpreting the Embedding Space) is provided above each subplot and measures the approximate deviation from a purely linear model. An error below 4\% predicts an approximately linear embedding line.  
}
\label{fig:reproducibility}
\end{suppfigure}
The line-like organization of our embedding space is a striking feature. Is this the result of chance, or is this a robust feature that may be reproduced reliably?

We explored this question by repeating our experiments, varying each the random seed used to generate the training images, as well as the random seed used to initialize the model perception network's weights. We show all the embeddings we obtained in Fig.~\ref{fig:reproducibility}. Each time we measured how line-like are the embeddings and we report the deviation from an exact line as a percent error below each embedding. We found that the deviations from a perfect line are very small, and most look perfectly linear with a few exceptions where we see slight kinks in the line.

\subsection{Restricting Dataset Variability}
\label{sec-jitter}
In our main experiment the arrangement of the objects in the scene varied randomly between {\em put}, {\em take} and {\em shake} actions. The size and contrast were varied as well. This was because we did not wish to presume that the agent (a child) playing with the objects would have to be careful with their motions. Furthermore we did not wish to presume that lighting conditions, and thus image contrast, and object pose, and thus their apparent size, would be preserved during the play session. However, one may suspect that scene randomness could help the model abstract the concept of ``number'' without being distracted by other factors such as object placement, contrast and size.

We explored the effect of randomness by modifying the process that generates data for Model B. In dataset B, object properties (area, intensity) are completely randomized before and after an action (Fig.~\ref{fig:main_data}B). We thus constructed a new dataset (Fig.~\ref{fig:jitter_dataset}), where we restricted the randomness before and after an action by reducing the amount of change in an object's area and intensity to a small amount of jitter. However, we still randomize object position, which we find is fundamental to learning a generalizable model of numerosity. We find that even after reducing object variation, the model has learned has the same properties as Model B (Fig.~\ref{fig:jitter_embedding}). However, learning is more sensitive to the initial seed (Fig.~\ref{fig:jitter_reproductibility}). We refer to this dataset as the {\em jitter dataset} and model's trained by this dataset as {\em Jitter Models}.

\begin{suppfigure}[h]
\begin{center}
    \renewcommand{\thesubsuppfigure}{\Alph{subsuppfigure}}
    \begin{subsuppfigure}{0.95\textwidth}
        \includegraphics[width=1\linewidth]{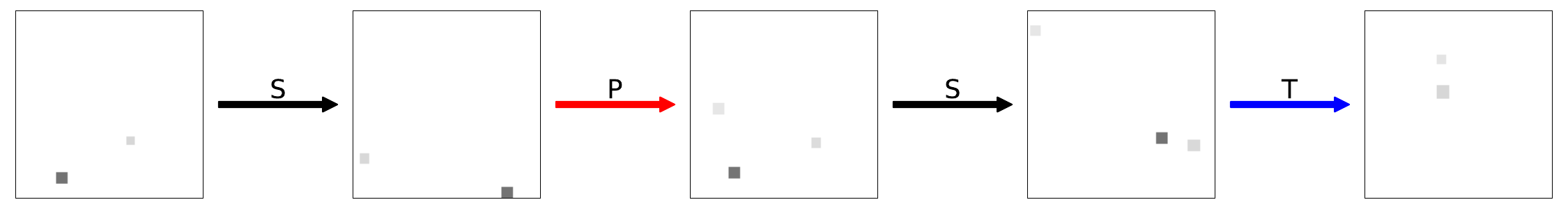}
    \end{subsuppfigure}
\end{center}
\linespread{1} \caption{\textbf{Jitter Datasets.} In Jitter Datasets, we restrict the change in size and contrast an object may undergo due to an action. After each action, the size (diagonal) of an object will be allowed to jitter by up to 3 pixels and the contrast by $\pm 0.02\%$ of the maximum contrast. We find that these small perturbations in object representations are sufficient to recreate similar results to those seen with Model B.}
\label{fig:jitter_dataset}
\end{suppfigure}

\begin{suppfigure}[h]
\begin{center}
    \renewcommand{\thesubsuppfigure}{\Alph{subsuppfigure}}
    \begin{subsuppfigure}{0.81\textwidth}
        \includegraphics[width=1\linewidth, trim={0 .2cm 0 0}, clip]{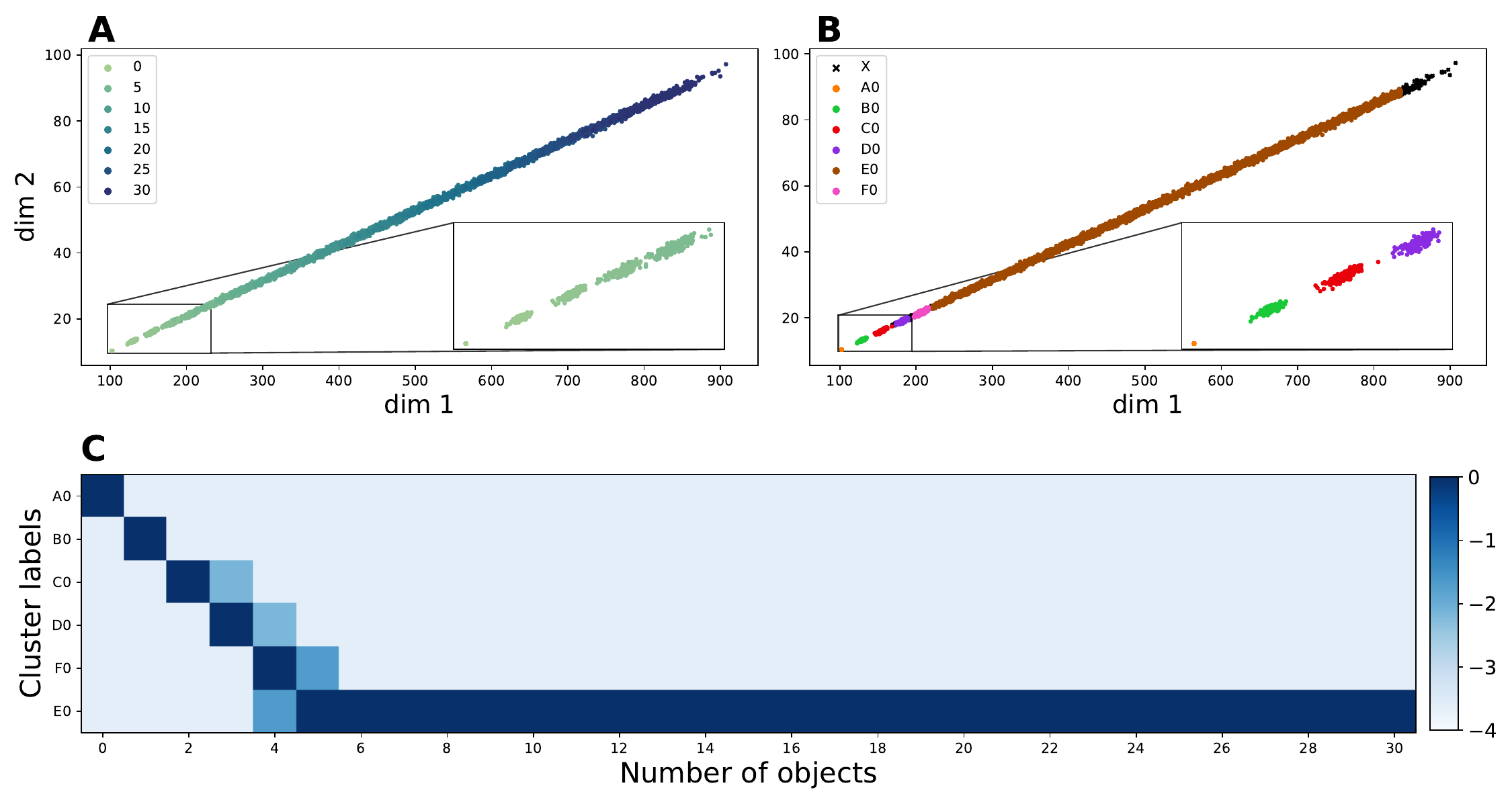}
    \end{subsuppfigure}
\end{center}
\linespread{1} \caption{\textbf{Properties of Jitter Models.} We find that the important properties of the Model B representation arise with Jitter Models. The model representations are linear, monotonic, with the early numbers easily separable. We set the minimum cluster size to 30 (HDBSCAN), in order to produce the most concise plots. Note the Jitter Model representations are more sensitive to minimum cluster size. }
\label{fig:jitter_embedding}
\end{suppfigure}

\begin{suppfigure}[h]
\begin{center}
    \renewcommand{\thesubsuppfigure}{\Alph{subsuppfigure}}
    \begin{subsuppfigure}{0.95\textwidth}
        \includegraphics[width=1\linewidth, trim={2cm 2cm 1.8cm 1cm}, clip ]{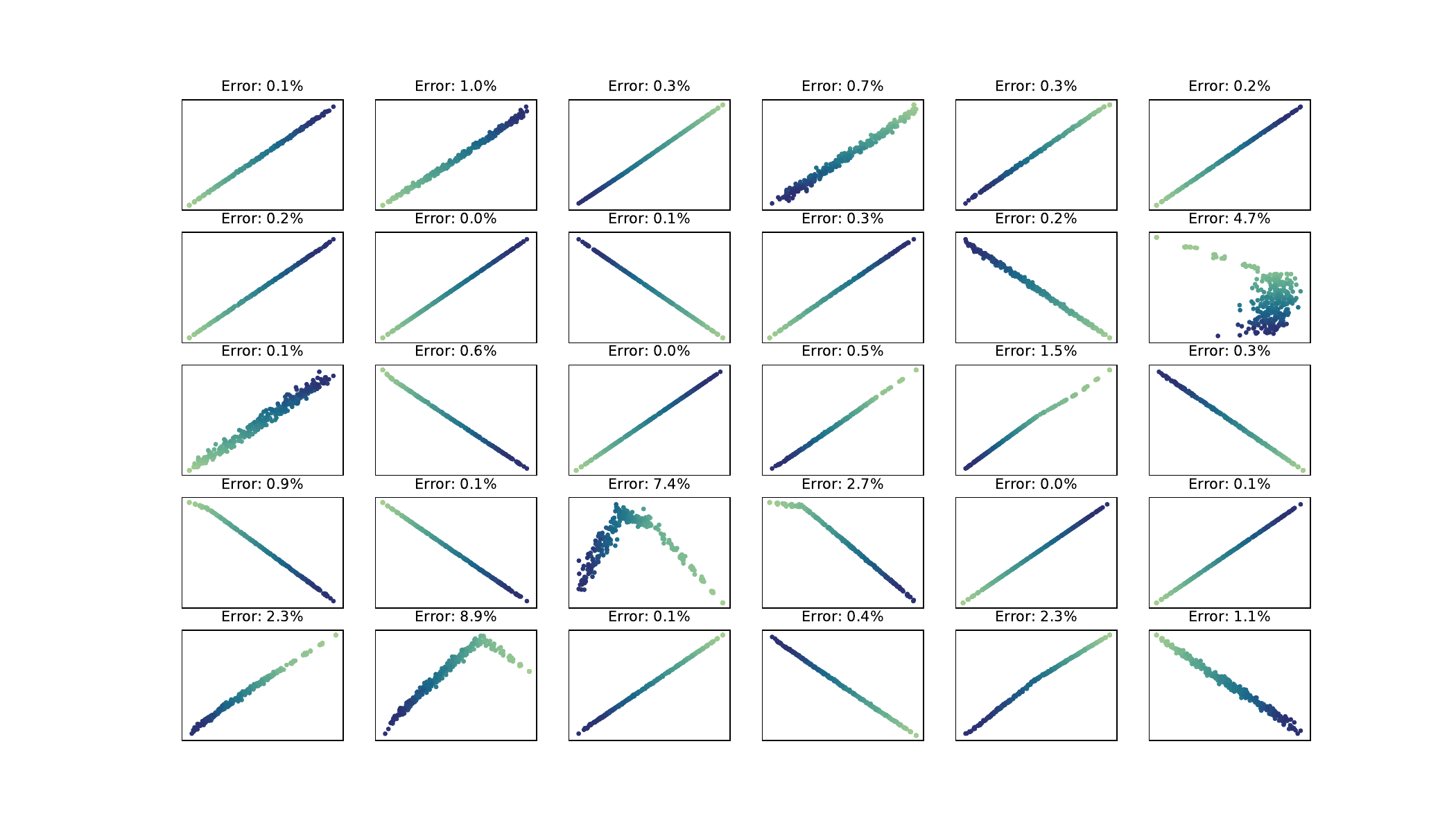}
    \end{subsuppfigure}
\end{center}
\linespread{1} \caption{\textbf{Reproducibility of Jitter Models.} We vary the initial seed to determine how reproducible the results are. We find model's trained with the jitter dataset learn mostly linear representations, however, certain seeds do result in large kinks. This indicates that visual variability between scenes will help the model learn the abstraction of number.}
\label{fig:jitter_reproductibility}
\end{suppfigure}


\subsection{Imprecise Action Sizes}
Will our model learn the abstraction of ``number'' even when the {\em put} and {\em take} actions will place or remove an unpredictable random number of objects?

We explored this question by randomizing the number of objects that each action affects in the range 0-3, as opposed to exactly 1 as in the main experiment. We capped the maximum number of objects to 3, like previous experiments. We find that while precise actions help in building distinct clusters in the subitization range, it is not necessary to retain the important properties of the generalizable number line. We refer to this dataset as the {\em imprecise actions dataset} (Fig.~\ref{fig:action_dataset}) and model's trained by this dataset as {\em Imprecise Action Models}. We find that all the properties of the original model retained (Fig.~\ref{fig:action_size_embedding}) and that the model is reproducible (Fig.~\ref{fig:action_size_reproductibility}).

\label{sec-action_size}
\begin{suppfigure}[h]
\begin{center}
    \renewcommand{\thesubsuppfigure}{\Alph{subsuppfigure}}
    \begin{subsuppfigure}{0.95\textwidth}
        \includegraphics[width=1\linewidth]{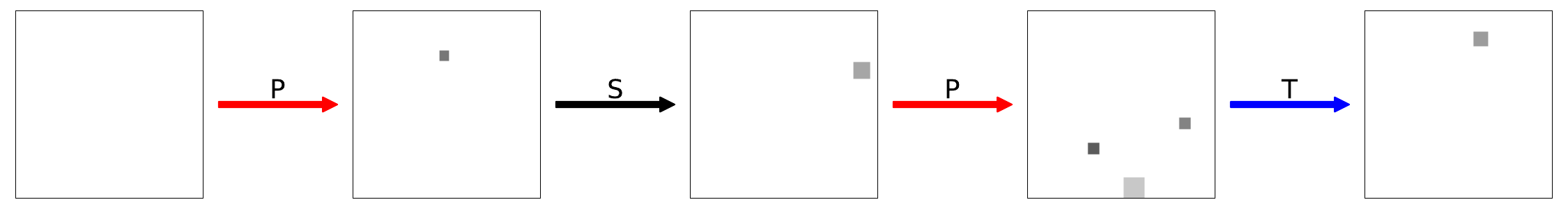}
    \end{subsuppfigure}
\end{center}
\linespread{1} \caption{\textbf{Imprecise Action Datasets.} In this dataset, we allow the number of objects taken or placed during an action to be 0-3 (limited by the number of objects in the visual scene). The maximum number of objects is still set to 3. This dataset mimics a situation in which the agent is imprecise with their actions and does not always select one object. The object's size and contrast are randomized between actions (like in dataset B). }
\label{fig:action_dataset}
\end{suppfigure}

\begin{suppfigure}[h]
\begin{center}
    \renewcommand{\thesubsuppfigure}{\Alph{subsuppfigure}}
    \begin{subsuppfigure}{0.81\textwidth}
        \includegraphics[width=1\linewidth, trim={0 0.2cm 0 0.0}, clip]{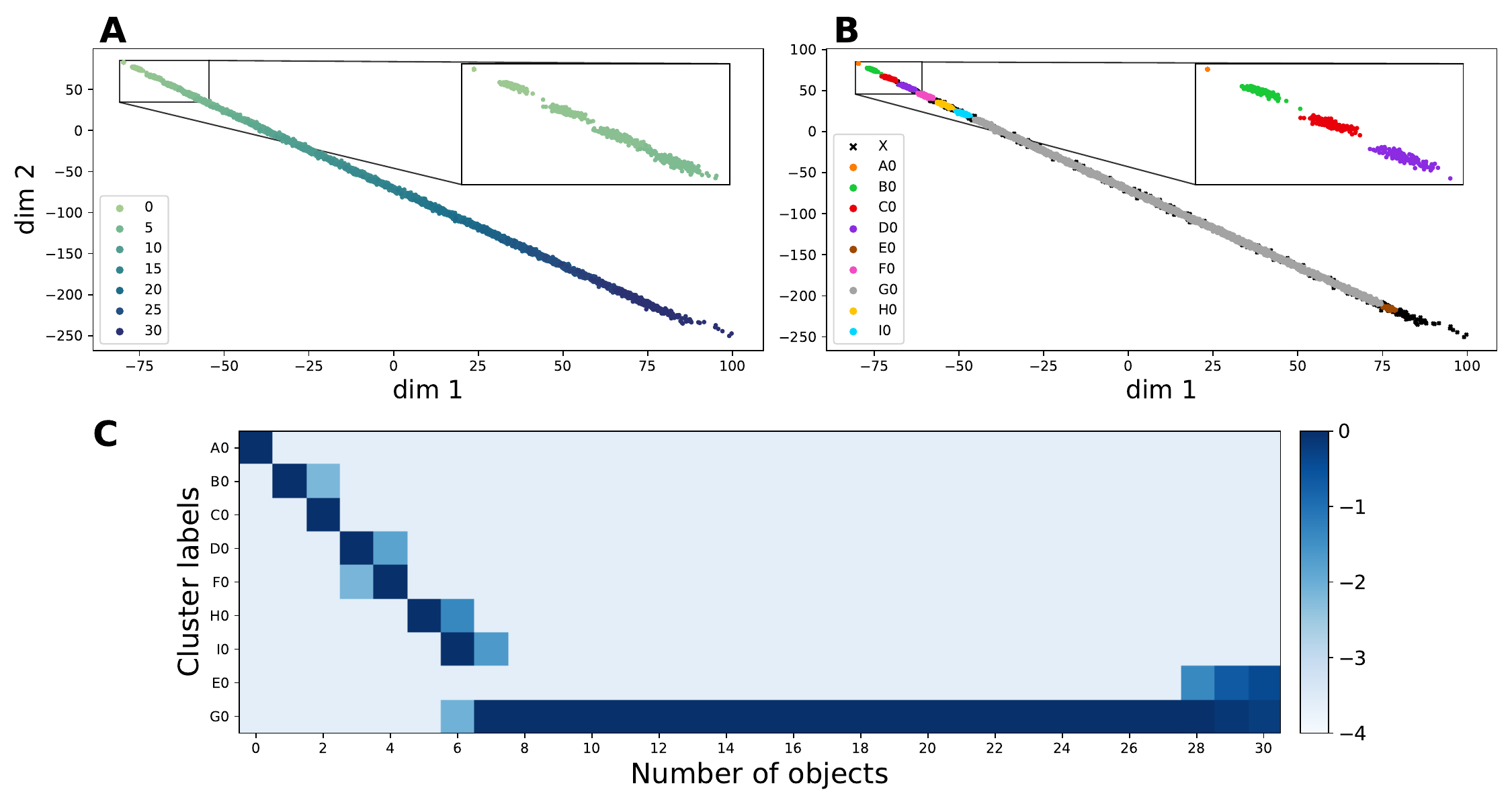}
    \end{subsuppfigure}
\end{center}
\linespread{1} \caption{\textbf{Properties of Imprecise Action Models.} We find that the important properties of the Model B representation arise with Imprecise Action Models. The model representations are linear, monotonic, with the early numbers easily separable. However, the separability of the early clusters is rougher than with precise action sizes. We set the minimum cluster size to 50 (HDBSCAN), in order to produce the most concise plots. Note the Imprecise Action Model representations are more sensitive to minimum cluster size.}
\label{fig:action_size_embedding}
\end{suppfigure}

\begin{suppfigure}[h]
\begin{center}
    \renewcommand{\thesubsuppfigure}{\Alph{subsuppfigure}}
    \begin{subsuppfigure}{0.95\textwidth}
        \includegraphics[width=1\linewidth, trim={2cm 2cm 1.8cm 1cm}, clip ]{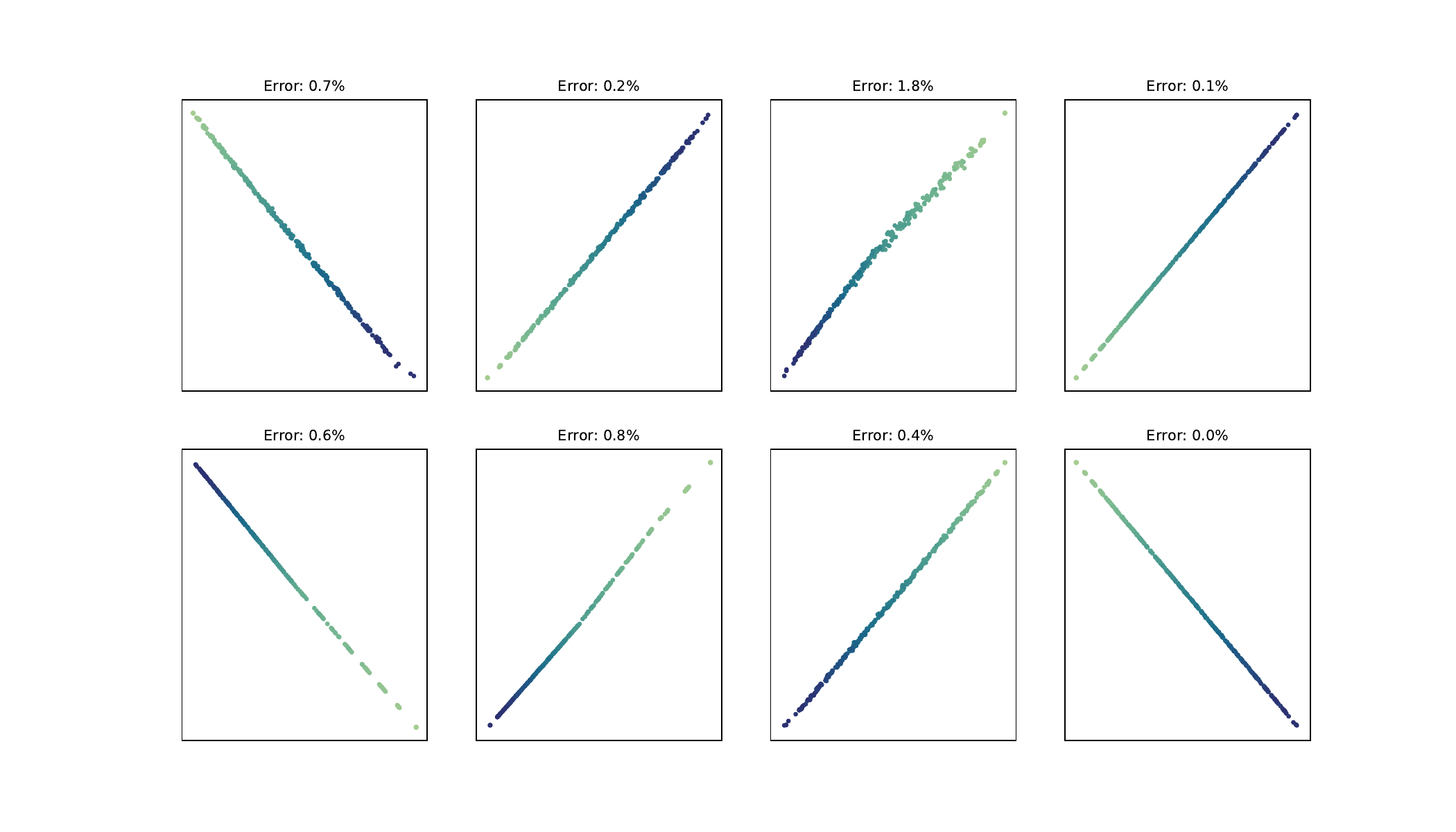}
    \end{subsuppfigure}
\end{center}
\linespread{1} \caption{\textbf{Reproducibility of Imprecise Action Models.} We vary the initial seed to determine how reproducible the results are. We find model's trained with the imprecise action sizes learn mostly linear representations.}
\label{fig:action_size_reproductibility}
\end{suppfigure}

\begin{suppfigure}[ht]
\section{Dataset Statistics}
\begin{center}
    \renewcommand{\thesubsuppfigure}{\Alph{subsuppfigure}}
    \begin{subsuppfigure}{0.46\textwidth}
        \includegraphics[width=1\linewidth]{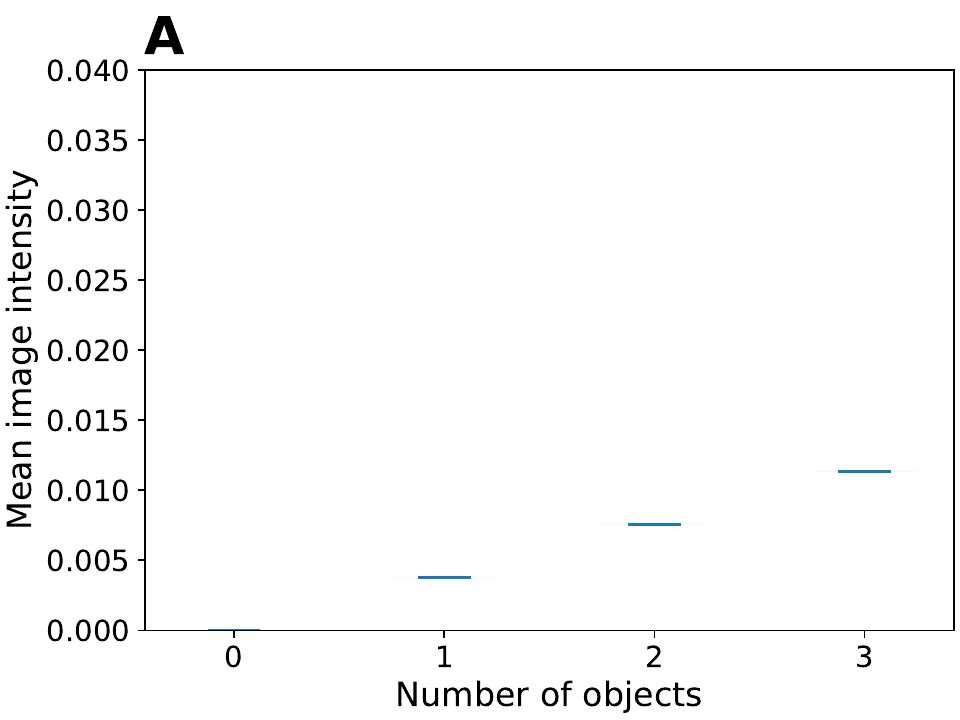}
    \label{fig:train_set1}
    \end{subsuppfigure}
    \begin{subsuppfigure}{0.46\textwidth}
        \includegraphics[width=1\linewidth]{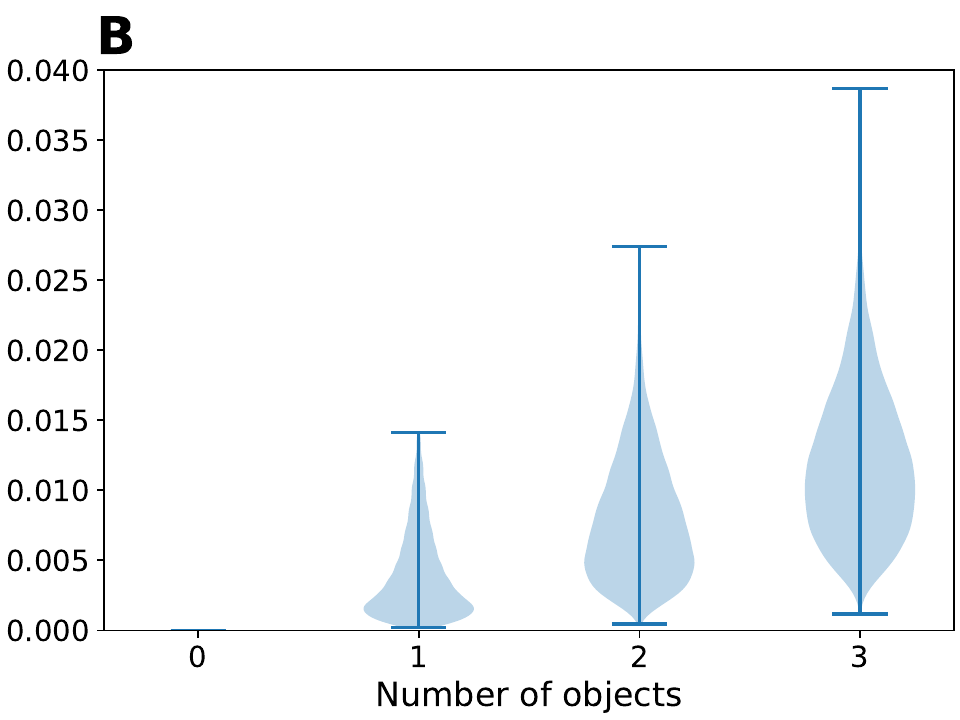}
    \label{fig:train_set2}
    \end{subsuppfigure}
    \begin{subsuppfigure}{0.46\textwidth}
        \includegraphics[width=1\linewidth]{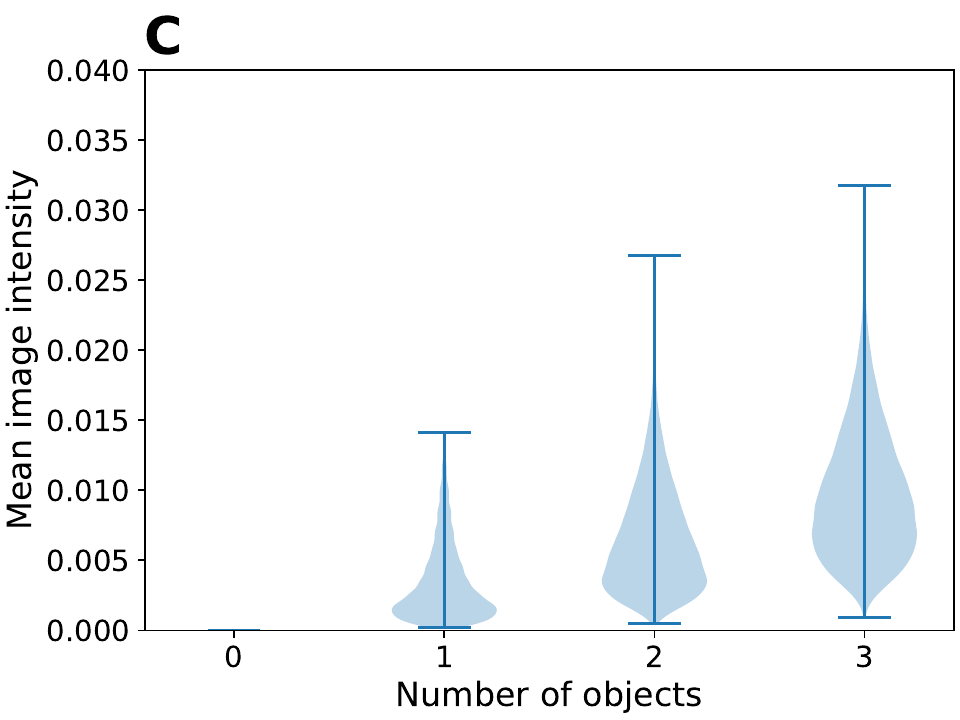}
    \label{fig:train_set3}
    \end{subsuppfigure}
    \begin{subsuppfigure}{0.46\textwidth}
    \includegraphics[width=1\linewidth]{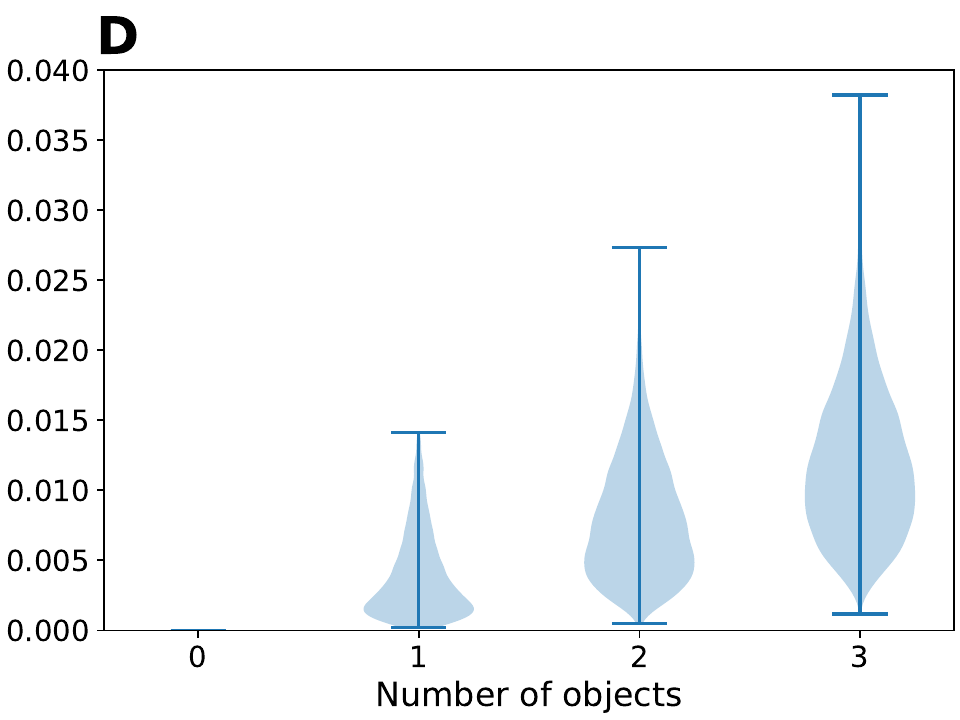}
    \label{fig:train_set4}
    \end{subsuppfigure}
\end{center}
\linespread{1} \caption{\textbf{Training set statistics.} \textbf{(A)} In dataset A (Fig.~\ref{fig:main_data}A) objects have the same size and contrast. Thus, the number of objects predicts the mean image intensity and vice-versa. \textbf{(B)} Objects in dataset B (Fig.~\ref{fig:main_data}B) have variable sizes and variable contrast, thus mean image intensity is not sufficient to predict the number of objects.
\textbf{(C)} Objects in the jitter datasets (Fig.~\ref{fig:jitter_dataset}) have a restricted, but variable size and contrast. We see the image statistics are similar to that of dataset B, but have a smaller amount of variability. \textbf{(D)} Objects in the imprecise actions datasets (Fig.~\ref{fig:jitter_dataset}) have random numbers of objects manipulated in an action. We see the image statistics are effectively the same as that of dataset B.}
\label{fig:dataset_statistics}
\end{suppfigure}

\begin{suppfigure}[t]
\section{Network Details}
\begin{center}
    \includegraphics[width=0.9\linewidth]{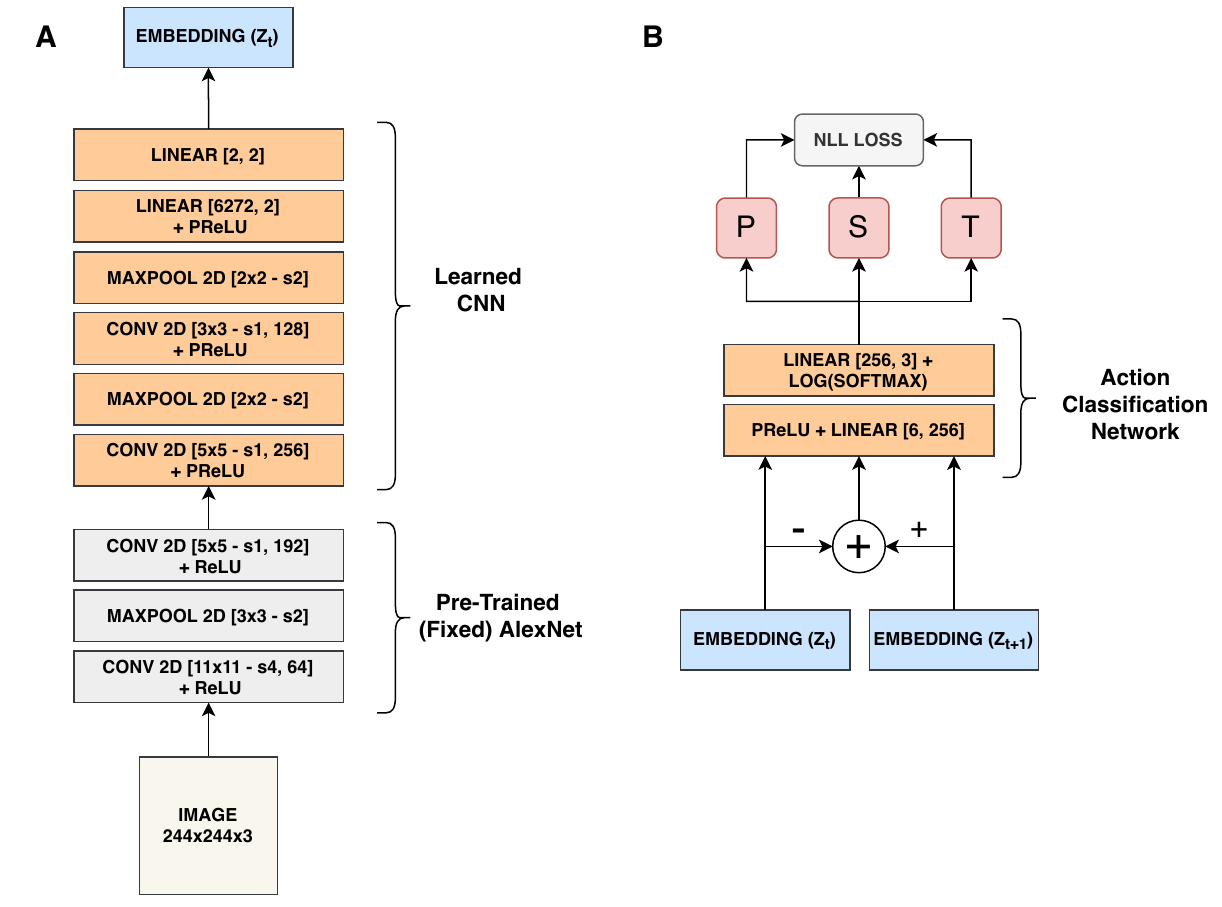}
\end{center}
\linespread{1} \caption{\textbf{Detailed diagram of the network structure.} \\\textbf{(A)} The feature extraction / embedding network. The gray layers are pre-trained on ImageNet~\cite{deng2009imagenet, krizhevsky2012imagenet} and remain fixed throughout the course of training. The orange layers are randomly seeded and trained simultaneously with the classifier in (B). The details of the layer are described within the brackets. For example, [11x11 - s4, 64] is an 11x11 kernel with a stride of 4 and 64 filters. During a training step, the embedding network accepts an image ($x_t$) of the visual scene and generates a lower-dimensional feature embedding ($z_{t}$) of the visual scene. An action: (P), (T), or (S) modifies the visual scene and the ``after" image ($x_{t+1}$) is passed through the embedding network as well. The outputs of the embedding network, ($z_{t}$) and ($z_{t+1}$) are treated as inputs to the action classification network. The shared embedding network is trained together with the classifier (B), in a Siamese configuration. \textbf{(B)} The action classification network is a 2-layer classifier network and is composed of two fully connected layers with a log-softmax transformation on the output. The input is the representation of the visual scene before and after an action is performed. The negative log-likelihood (NLL) loss function is used to train both the action classification network and the embedding network simultaneously. An overview of the entire training paradigm is shown in Figure \ref{fig:network}.}
\label{fig:network_details}
\end{suppfigure}
